\documentclass[review]{elsarticle}
\usepackage{subfigure,color}
\usepackage{lineno,hyperref}
\modulolinenumbers[5]

\journal{Journal of \LaTeX\ Templates}

\begin{document}

\begin{frontmatter}

\title{The characteristics of cycle-nodes-ratio and its application to network classification}
\author{Wenjun Zhang$^{1, 3}$, Wei Li$^{1, 2*}$ and Weibing Deng$^{1*}$  }
\address{$^1$ Key Laboratory of Quark and Lepton Physics (MOE) and Institute of Particle Physics,  Central China Normal University, Wuhan 430079, China}
\address{$^2$Max-Planck Institute for Mathematics in the Sciences, Inselstrasse 22-26, D-04103 Leipzig, Germany}
\address{$^3$School of Medical Information Engineering, Anhui University of Chinese Medicine, Hefei 230012, China}

\ead{liw@mail.ccnu.edu.cn, wdeng@mail.ccnu.edu.cn}

\begin{abstract}
Cycles, which can be found in many different kinds of networks, make the problems more intractable, especially when dealing with dynamical processes on networks. On the contrary, tree networks in which no cycle exists, are simplifications and usually allow for analyticity. There lacks a quantity, however, to tell the ratio of cycles which determines the extent of network being close to tree networks. Therefore we introduce the term Cycle Nodes Ratio (CNR) to describe the ratio of number of nodes belonging to cycles to the number of total nodes, and provide an algorithm to calculate CNR. CNR is studied in both network models and real networks. The CNR remains unchanged in different sized Erd\H os-R\' enyi (ER) networks with the same average degree, and increases with the average degree, which yields a critical turning point. The approximate analytical solutions of CNR in ER networks are given, which fits the simulations well. Furthermore, the difference between CNR and two-core ratio (TCR) is analyzed. The critical phenomenon is explored by analysing the giant component of networks. We compare the CNR in network models and real networks, and find the latter is generally smaller. Combining the coarse-graining method can distinguish the CNR structure of networks with high average degree. The CNR is also applied to four different kinds of transportation networks and fungal networks, which give rise to different zones of effect. It is interesting to see that CNR is very useful in network recognition of machine learning.
\end{abstract}


\end{frontmatter}

\section{\label{sec:introduction}Introduction}
The study of network structures is a key point in the initial stage of network science. The cycle property as the basic property of graph has been studied extensively \cite{Solimano1982Graph}. The cycles play crucial roles in network spreading \cite{CantwellMessage,kuikka2018influence}, community detection \cite{kuikka2018influence}, and network control \cite{rozum2018identifying}, etc. Bianconi \textit{et al} \cite{Bianconi2003} studied the statistic of cycles in Barab{\'a}si-Albert (BA) \cite{barabasi1999emergence}  scale-free networks and obtained: $\langle N_h(t)\rangle=[(m/2)\log(t)]^h[1+O(\zeta^{-1})]\sim[(m/2)\log(t)]^h$, where $\zeta=m/(2\log (t))$, $N_h(t)$ represents the number of loops with length $h$, $m$ represents the number of added edges at every step in BA model, and $t$ is size of network. Noh \cite{Noh2008Loop} calculated the distributions of cycles in both uncorrelated networks and correlated ones. At nearly the same time the study of the tree property in complex networks began to prevail. Gu \textit{et al} \cite{Gu2010A} incorporated small-world effect and scale-free property into a tree-like network model. Wechsatol \textit{et al} \cite{Wechsatol2005} found the cycles can enhance the robustness of the network. And it was paid attention again recently to study the effects of cycles \cite{Wylie2012Linked,fan2019towards,Xu2017Optimal,shi2019totally,Wu2018Bridges}. Wylie \textit{et al} \cite{Wylie2012Linked} found the switch integration in biological networks is controlled by the cycles but not the network hubs. Fan \textit{et al} \cite{fan2019towards} studied the properties of loops in network models, based on which the node's importance can be quantified. Given the proportion of minimum feedback vertex set, Zhou \textit{et al} \cite{zhou2013spin,Xu2017Optimal} proposed an algorithm to search for the minimum number of feedback edges, and they concluded the feedback vertices are important for the structural and dynamical properties of directed networks. Shi \textit{et al} \cite{shi2019totally} proposed to search for linearly independent cycles of a network, which was used in studying the cyclic structures in small-world networks. Wu \textit{et al} \cite{Wu2018Bridges} discussed the bridges in complex networks, and proposed the concept of bridgeness to represent edge centrality. Lately, there has been great interest in network classification by using the method of machine learning. Kantarc \textit{et al} \cite{kantarci2013classification} classified the networks by the topological quantities. They considered 14 different topological measures to characterize the networks, and found that the most favorable measures are density, modularity, average degree, and transitivity (global clustering coefficient). Niousha \textit{et al} \cite{attar2017classification} also proposed the method of network classification based on the supervised learning and distance-based classification algorithm, which has high classification accuracy, noise tolerance, and computation efficiency.

Although the structure of cycles is important and some qualities can describe the richness of cycles. But there is no more intuitive quality, to our best knowledge, to tell whether a given network tends to be tree-like or not. We have studied the cycle properties in Ref.~\cite{zhang2018statistical}, and found that it is crucial for transportation whether the ports are located along with cycles or not in world marine networks. Here we introduce the Cycle Nodes Ratio (CNR) of a given network, defined as the ratio of nodes belonging to cycles to the number of total nodes. The algorithm of calculating the number of cycles is hard in networks. Here we propose an algorithm, named CDFS, based on the Depth First Search (DFS) \cite{tarjan1972depth} algorithm, to calculate the CNR for the network of interest. The LDFS has low computation complexity $O(n+m)$, where $n$ and $m$ are the numbers of nodes and edges, respectively. We calculate the CNR of three basic random networks and identify critical transition points in Erd\H os-R\' enyi (ER) \cite{erdHos1960evolution} networks, as well as the underlying mechanism. The approximate analytical solution of the  CNR in ER networks is given by calculating the giant component in ER networks. Otherwise, the CNR is 1 in both BA networks and Watts-Strogatz (WS) networks \cite{watts1998collective} with average degree no less than 4. In this process, we proved the key difference between CNR and two-core ratio (TCR) in WS network and fungal growth network. We also calculate the CNR in real networks by using different data sets, which are slightly lower than the counterparts of ER networks with the same average degree. To judge the CNR properties of networks in which the smallest degree is greater than 1, we combine the improved spectral coarse-graining (ISCG) \cite{zhou2017improved} to test the CNR of the macroscopic property of this kind of networks. Last but not least, we compare the CNR in different kinds of networks, and consider the CNR, density, modularity \cite{newman2004fast}, average degree, and global clustering coefficient \cite{wasserman1994social}, as the features to be classified by K-Nearest Neighbor (KNN) algorithm \cite{keller1985fuzzy}. We have found that CNR is the most important feature for network classification.

The rest of this paper is organized as follows: the concept of CNR and the corresponding algorithm are given in Sec.~\ref{sec:cyclenodesO}. Sec.~\ref{sec:CNRinnetworks} discusses the CNR properties in random networks and the approximate analytical solution of CNR in ER networks. The CNR is studied for different kinds of complex networks in Sec.~\ref{sec:CNRinclassification}, as well as its significance in classification of networks by machine learning. The last section contains the conclusion including the applications and the improvement of CNR.

\section{\label{sec:cyclenodesO}The definition of cycle nodes ratio and its algorithm}
\subsection{Cycle nodes ratio}
\begin{figure}[htbp]
\begin{tabular}{cc}
    \subfigure[Cycle in direct graph.]{\label{fig:subfig:a}
\includegraphics[width=0.42\textwidth]{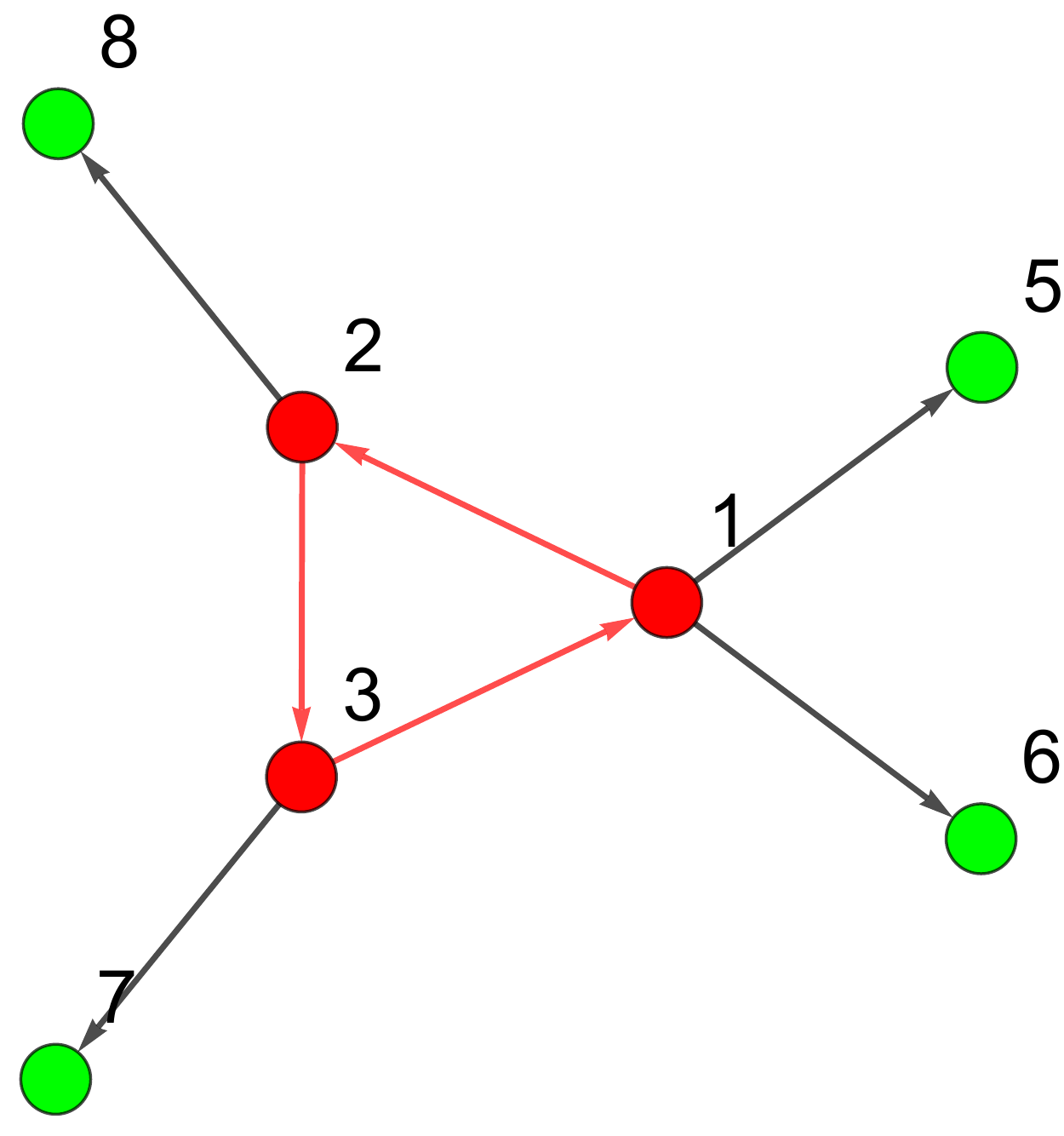}}
&
  \subfigure[Cycle in undirect graph.]{\label{fig:subfig:b}
\includegraphics[width=0.42\textwidth]{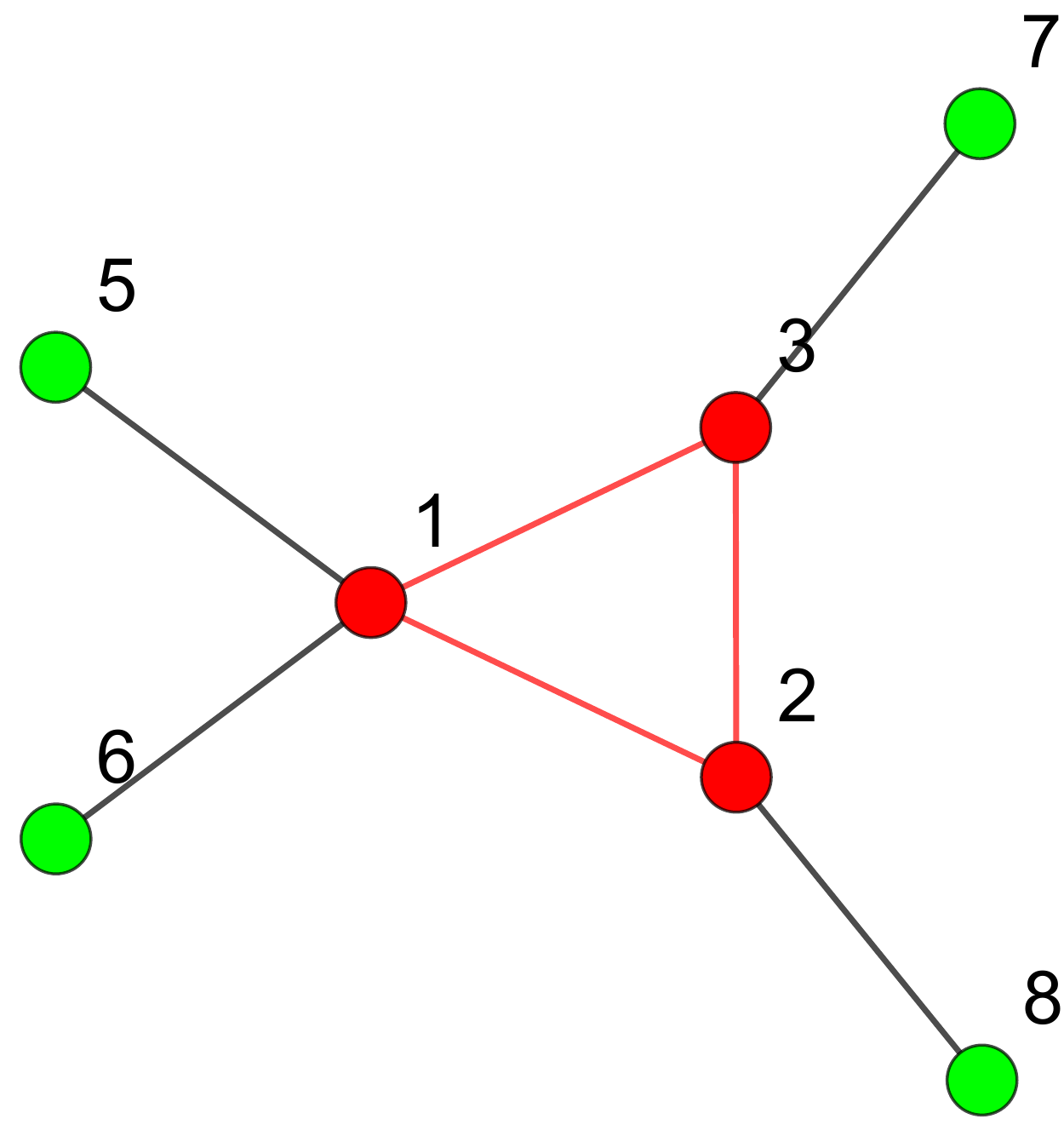}}
\end{tabular}
\caption{\label{fig:cycle}(Colour online) The graphical representation of cycle in directed network (a) and undirected network (b) where the red lines and nodes constitute cycle.}
\end{figure}

As we know, the cycle in a network is defined as: a cycle is an elementary chain which returns to the initial node \cite{holzmann1972tree}, as shown in Fig.~\ref{fig:cycle}. And we introduce the definition of the cycle nodes as:

\textbf{Cycle node:} the node sites along one or more cycle(s) path in a network.

\begin{figure}[htp]
\centering
\includegraphics[width=.7\textwidth]{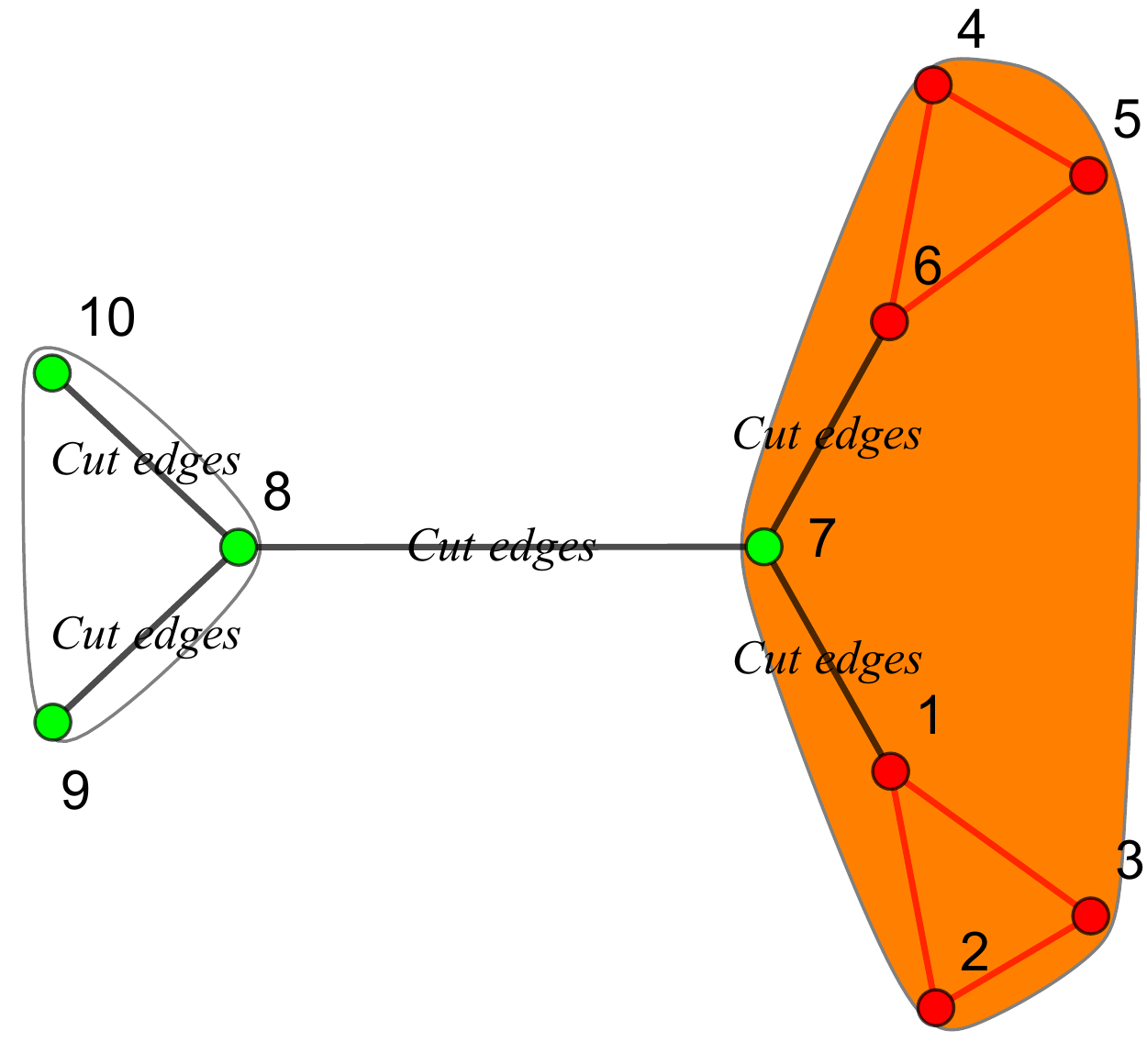}
\caption{(Colour online) A graph with 10 nodes and 9 edges, which has two cycles $1\leftrightarrow 2\leftrightarrow 3\leftrightarrow 1$ and $4\leftrightarrow 5\leftrightarrow 6\leftrightarrow 4$. So nodes 1, 2, 3, 4, 5, 6 are cycle nodes. The black line is cut edges. Otherwise, the orange region specifies 2-core nodes, which mean that the node 7 is 2-core but not a cycle node.
}
\label{fig:cyclenodes}
\end{figure}

An example of a cycle nodes is shown in Fig.~\ref{fig:cyclenodes}. The concept of cycle node is different from the concept of k-core, which is defined as the remaining nodes by recursively deleting the leaf nodes \cite{pittel1996sudden}, We refer to nodes similar to node 7 as cycle connect node which is not cycle node but can connect different cycles. The proportion of cycle nodes in a network can represent the characteristics of how far a network is away from tree networks in which no cycle exists. The CNR can be defined as:

\textbf{Cycle nodes ratio:} Suppose for an undirected, connected graph $G(V, E)$ with $n$ vertices and $m$ edges, the number of cycle nodes is $n_l$. The cycle nodes ratio is $R=n_l/n$, where $n$ is the number of nodes in $G(V, E)$.

\begin{figure}[htbp]
\centering
  \begin{tabular}{cc}
  \subfigure[Cycle with 50 nodes.]{\label{fig:cycleandtree:a}
    \includegraphics[width=0.42\textwidth]{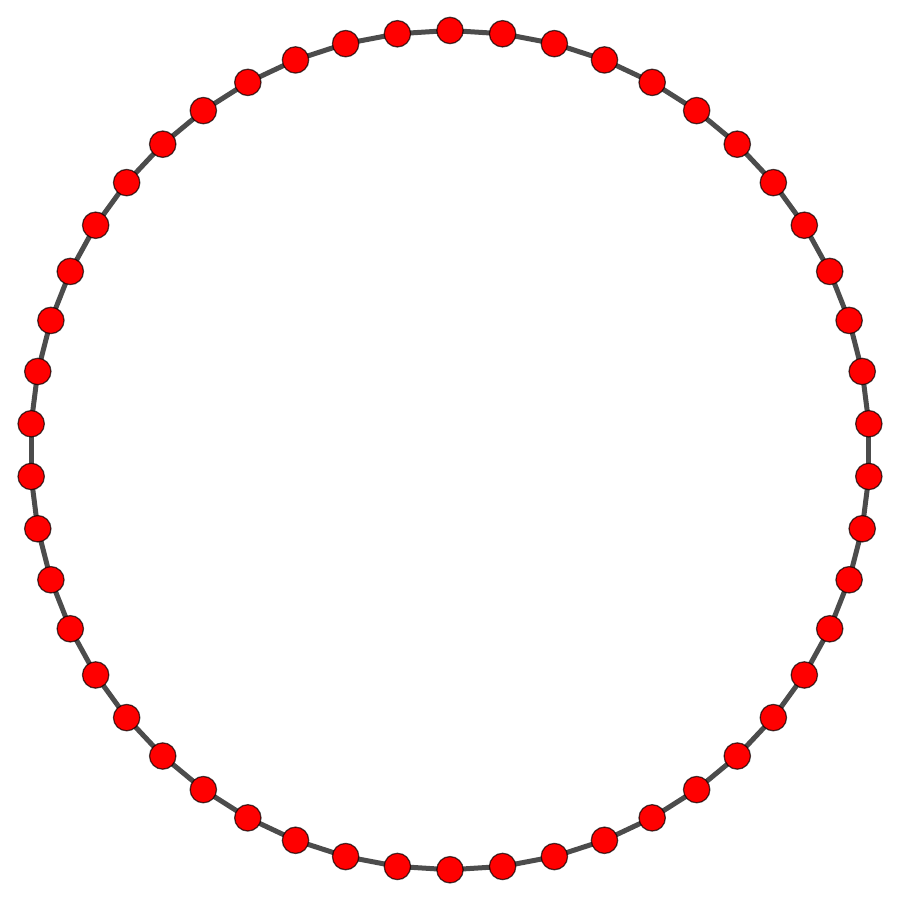}} &
     \subfigure[Tree with 50 nodes.]{\label{fig:cycleandtree:b}
    \includegraphics[width=0.42\textwidth]{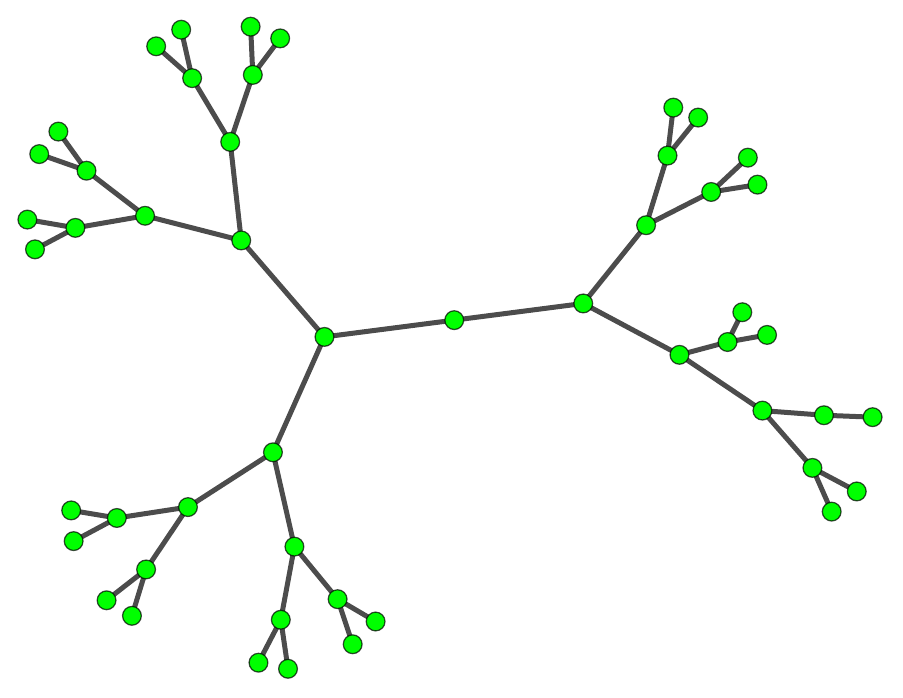}}
  \end{tabular}
  \caption{(Colour online) (a) A cycle graph with 50 nodes and 50 edges in which CNR is 1. (b) A tree graph with 50 nodes and 49 edges in which CNR is 0.}
  \label{fig:cycle_and_tree}
\end{figure}

The CNR is 1 in cycle graph and 0 in tree (star) graph as shown in Fig.~\ref{fig:cycle_and_tree}. Clearly, the normalization is satisfied.

\subsection{\label{sec:algorithm}The algorithm of calculating CNR}
The statistics of the cycles in complex networks is a hard job \cite{Bianconi2003}. It is almost impossible to judge whether a node is cycle node by calculating the number of cycles in a given network. But we find that a node is cycle node if there is no-cut edge connecting to it. The cut edge is defined in graph theory \cite{fleischner1990eulerian} as:

\textbf{Cut edges:} In a graph, if removing a given edge causes the addition of one connected component of this graph, then this edge is called cut edge, which also known as bridge (Shown in Fig.~\ref{fig:cyclenodes}).

\begin{figure}[htbp]
\centering
  \begin{tabular}{cc}
  \subfigure[The graph with 13 nodes.]{\label{fig:DFS:a}
    \includegraphics[width=.47\textwidth]{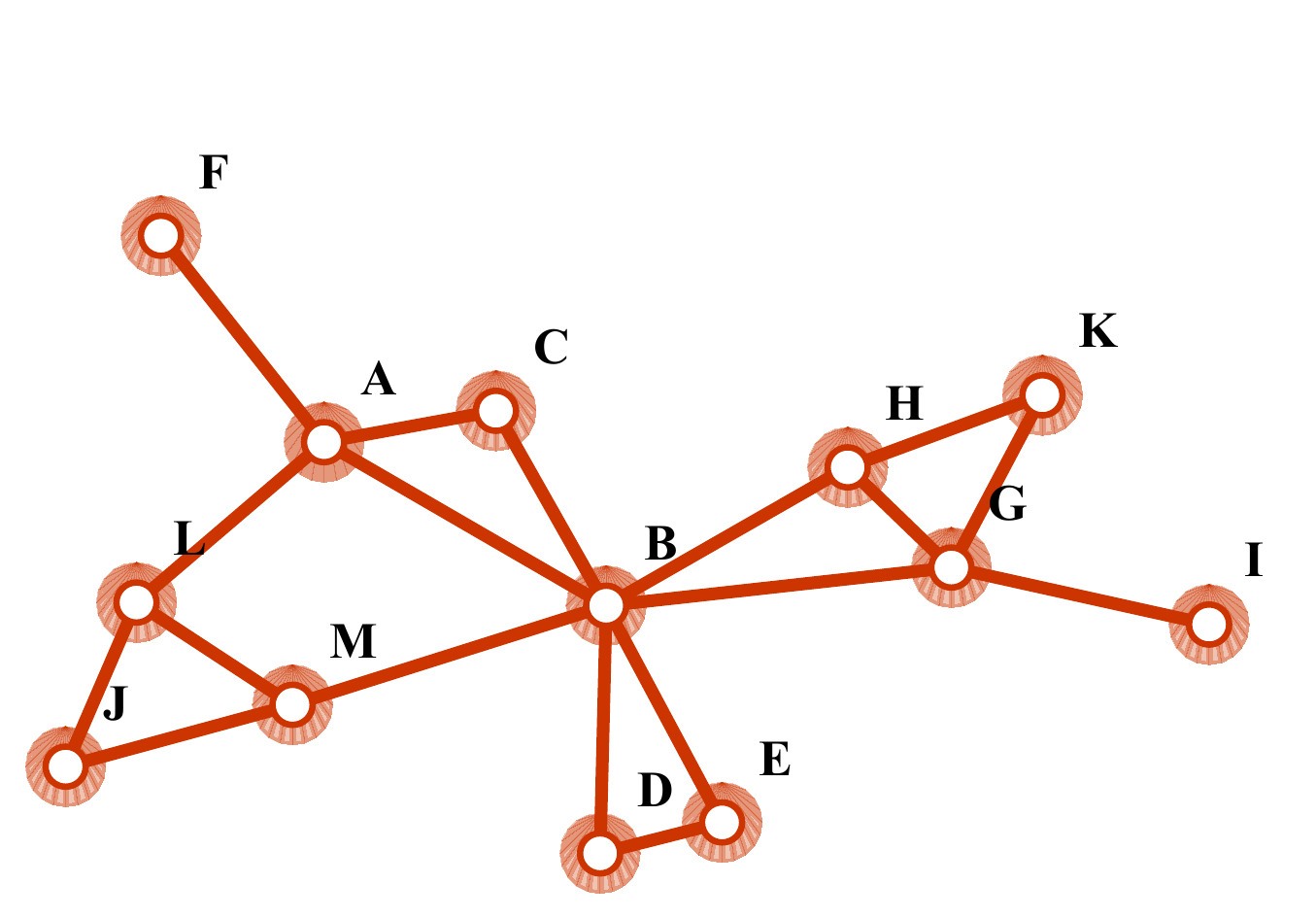}} &
     \subfigure[The DFS result of Fig.~\ref{fig:DFS:a}.]{\label{fig:DFS:b}
    \includegraphics[width=.35\textwidth]{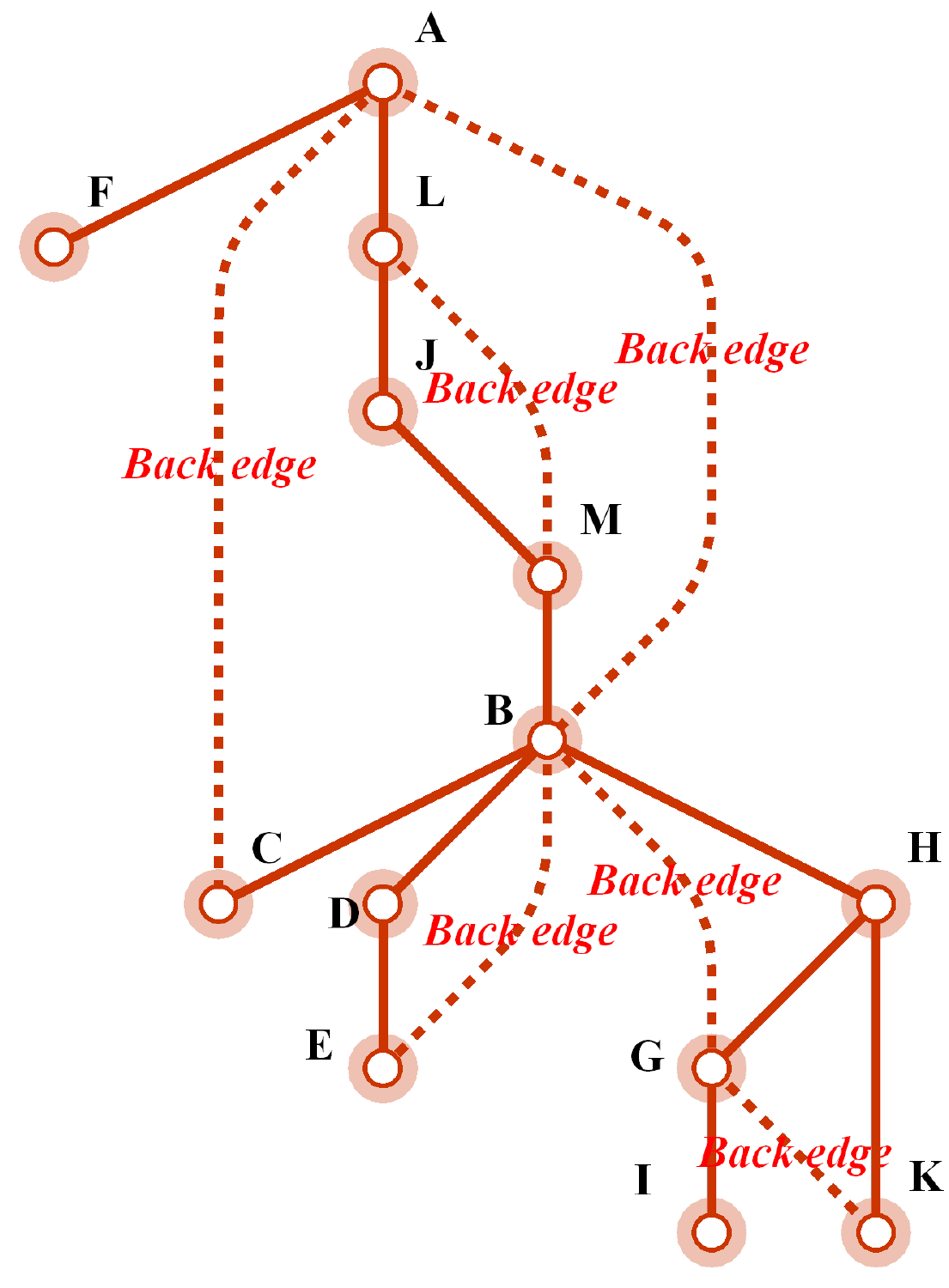}}
  \end{tabular}
  \caption{(Colour online) (a) A graph with 13 nodes and 18 edges. (b) The graph representation of DFS search tree. The dashed line are back edge.}
  \label{fig:tarjan}
\end{figure}

The edge is simply called no-cut edge if it is not cut edge, and the cut edges can be recognized by Tarjan algorithm \cite{tarjan1972depth}. We here introduce some basic concepts of Tarjan algorithm:
\begin{enumerate}
\item Depth First Search (DFS) search tree: We can get the DFS search tree from graph ergodic by DFS, shown in Fig.~\ref{fig:DFS:b}.
\item Tree edges: In the DFS process, the passing edges of visiting the nodes which are not visited, like the edges not labeled in Fig.~\ref{fig:DFS:b}.
\item Back edges: In the DFS process, the passing edges of visiting the nodes which are visited, like the edges labeled in Fig.~\ref{fig:DFS:b}.
\end{enumerate}

For a node $u$, we need to record two values that are $dfn[u]$ which represents the order of node $u$ in the DFS process, and $low[u]$ which represents the node $u$ that can retrospect the earliest node and not pass through its father node. The $dfn[u]$ and $low[u]$ are added  1 in the each step of DFS process, then we recalculate the $low[u]$ from the root node by following the rule that
\begin{displaymath}
low[u]=\left\{
\begin{array}{ll}
\textrm{min}\{ low[u], low[v]\} & (u,v) \textrm{ is tree edge},\\
\textrm{min}\{ low[u], dfn[v]\} & (u,v) \textrm{ is back edge}.\nonumber
\end{array} \right.
\end{displaymath}
The edge $(u,v)$ is cut edge if $low[v]>dfn[u]$. The DFS result of Fig.~\ref{fig:DFS:a} is shown in Table.~\ref{Tab:Tarjanresults}, the top line represents the node, and the second line and third line represent the $dfn$ and $low$ of the corresponding node respectively. Otherwise, the DFS just traverses the nodes which are in the same component as father node of this algorithm. So we can take one chosen node of each component as the father node of DFS for networks with multi-components. This algorithm is shown in Ref.~\cite{CNR2019}. Then we can get the corresponding cut edges. For a graph $G(V, E)$ with $n$ vertices and $m$ edges, the time complexity of this algorithm is $O(n+m)$ for adjacency list data or $O(n^2)$ for adjacency matrix data. This algorithm has high efficiency in classifying cycle nodes and non-cycle nodes in a network.
\begin{table}[!htp]
\centering
\caption{The DFS results of Fig.~\ref{fig:DFS:a}.}
\label{Tab:Tarjanresults}
\begin{tabular}{lccccccccccccc}
\hline
Node&A&B&C&D&E&F&G&H&I&J&K&L&M\\
\hline
$dfn[i]$&1&5&12&11&10&13&8&6&9&4&7&2&3\\
$low[i]$&1&1&1&5&5&13&5&5&9&2&5&1&1\\
\hline
\end{tabular}
\end{table}

\section{\label{sec:CNRinnetworks}The simulation and analytical results of CNR in networks}
\subsection{\label{CNRinmodel}The simulation results of CNR and its applications in basic random networks}
\begin{figure}[htbp]
\centering
  \begin{tabular}{cc}
  \subfigure[The CNR result in fixed average degree.]{\label{fig:CNRER:a}
    \includegraphics[width=.52\textwidth]{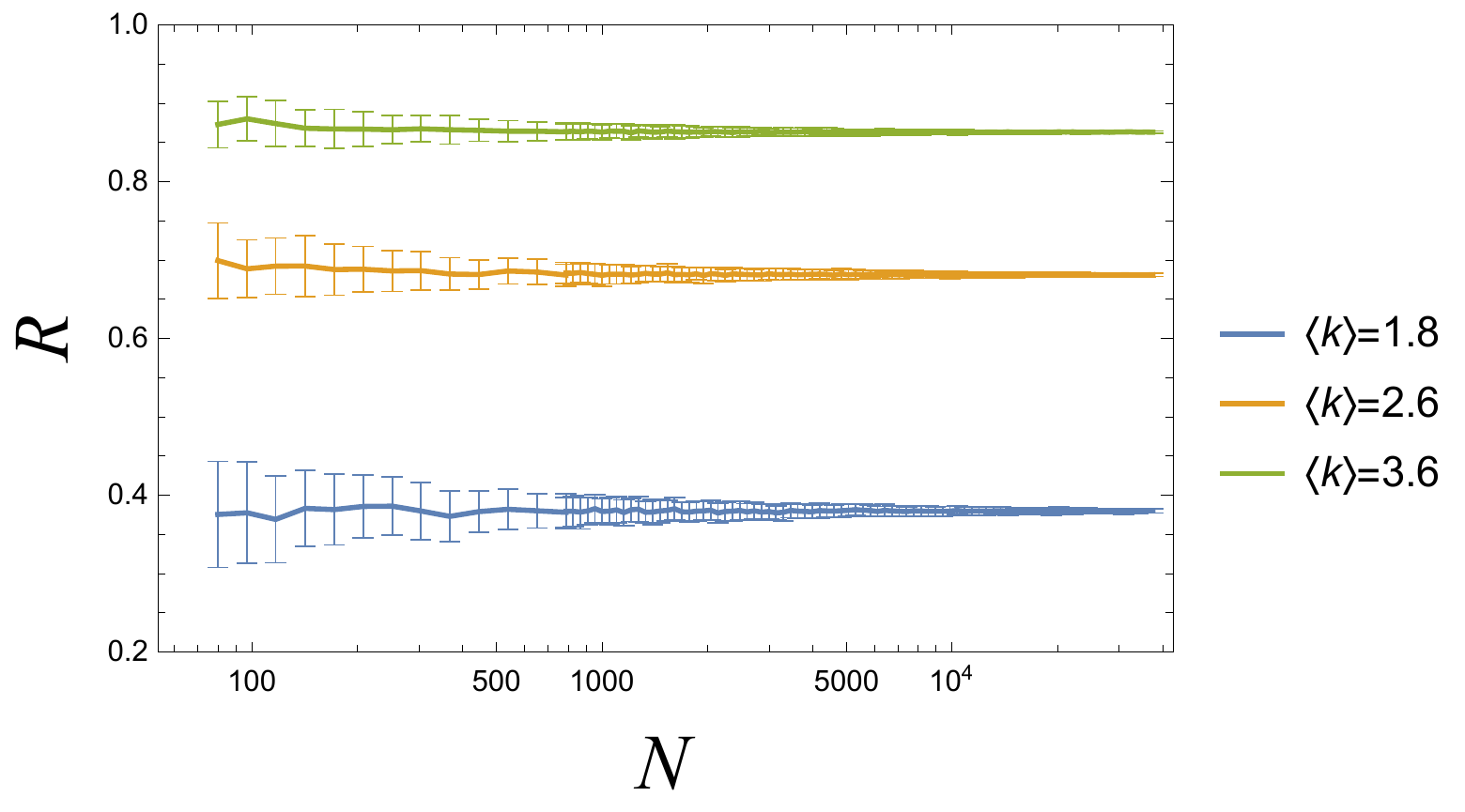}}&
    \subfigure[The CNR result in fixed networks size.]{\label{fig:CNRER:b}
    \includegraphics[width=.43\textwidth]{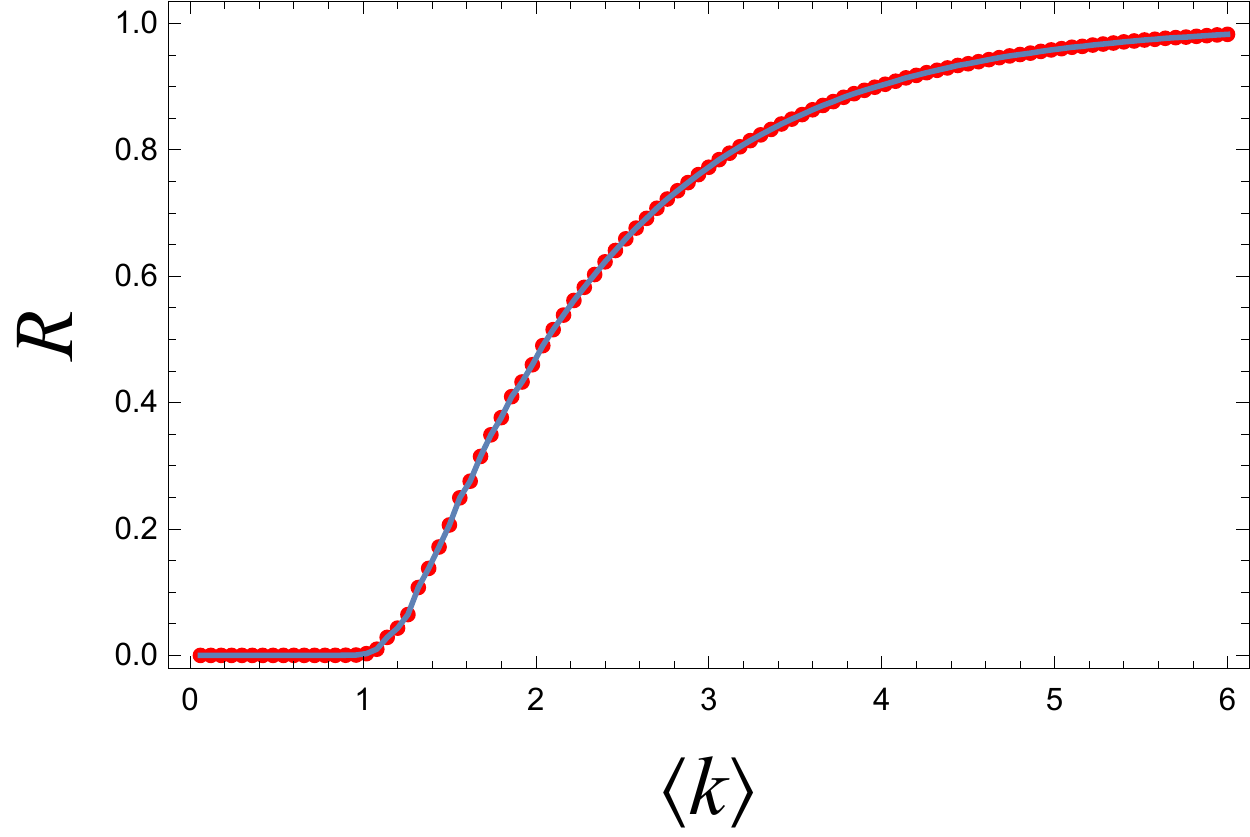}}
  \end{tabular}
  \caption{(Colour online) (a) The relation between CNR and the number of nodes in ER networks with fixed average degree of 1.8, 2.6, and 3.6. (b) The relation between CNR and average degree of ER networks with 10000 nodes. The red lines are fittings, and the error bars are too small to be visible. Moreover, $R$ represents the CNR, $k$ represents the average degree of networks, and $N$ represents the number of nodes in networks. (In this paper, all simulation results were averaged over 200 different realizations. And all error bar is the value of standard deviation.)}
  \label{fig:CNRinERmodel}
\end{figure}

To study the properties of CNR, we calculate its value in ER networks. In the first approach, we fix the average degree and change the size of networks, and in the second one, we fix the size of networks and change the average degree. The results are shown in Fig.~\ref{fig:CNRinERmodel}. The major findings are concluded as follows:
\begin{itemize}
\item The CNR of ER model remains constant in different sized networks if the average degree of ER networks is fixed.
\item If we fix the size of ER networks, the value of CNR will increase with the average degree of ER networks.
\item A critical turning point occurs when the average degree of ER networks is 1.
\end{itemize}

\begin{figure}[htbp]
\centering
  \begin{tabular}{cc}
  \subfigure[The CNR result in WS networks with fixed rewiring probability.]{\label{fig:CNRws:a}
    \includegraphics[width=.42\textwidth]{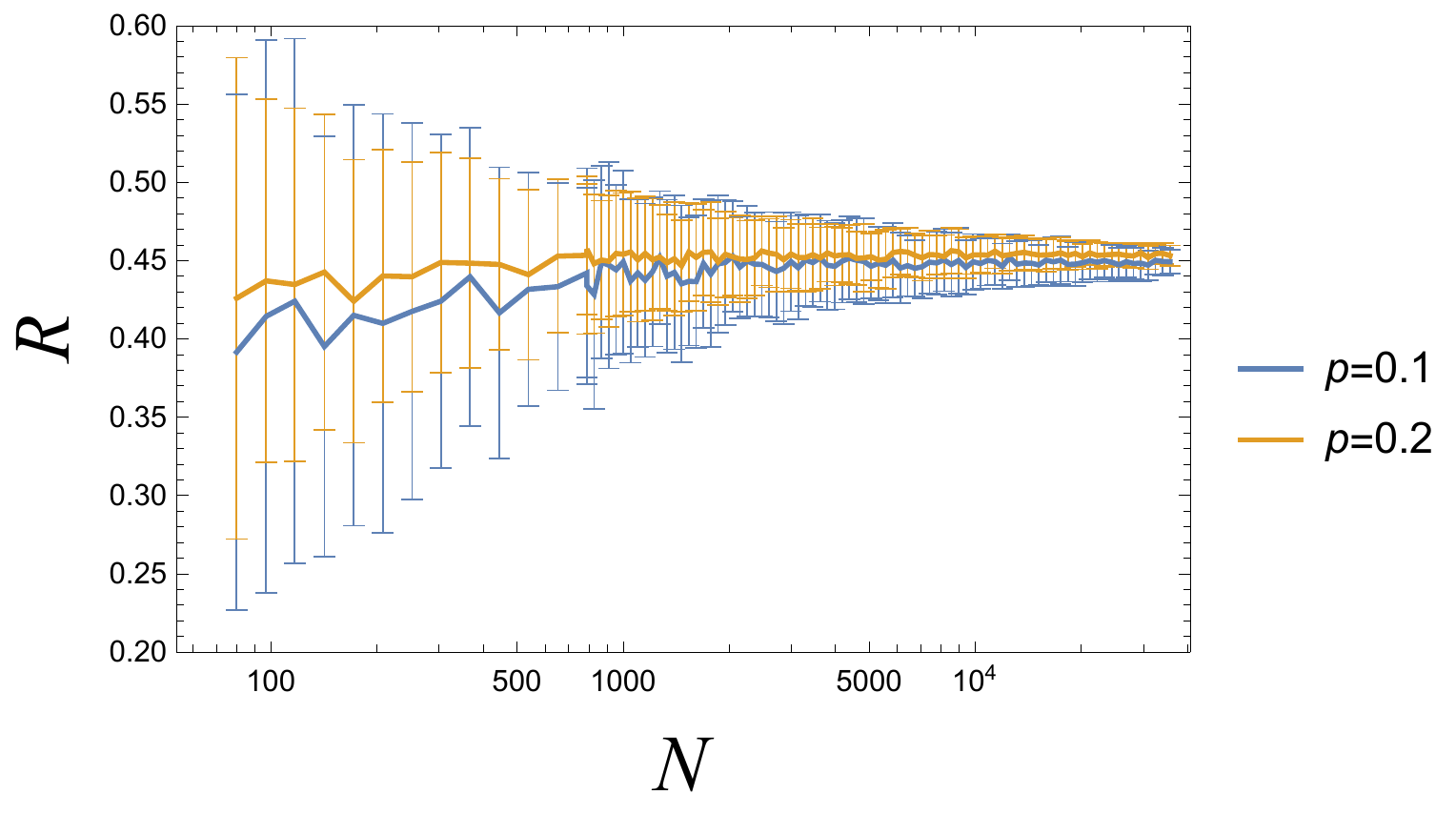}}&
    \subfigure[The CNR result in WS networks with fixed size.]{\label{fig:CNRws:b}
    \includegraphics[width=.53\textwidth]{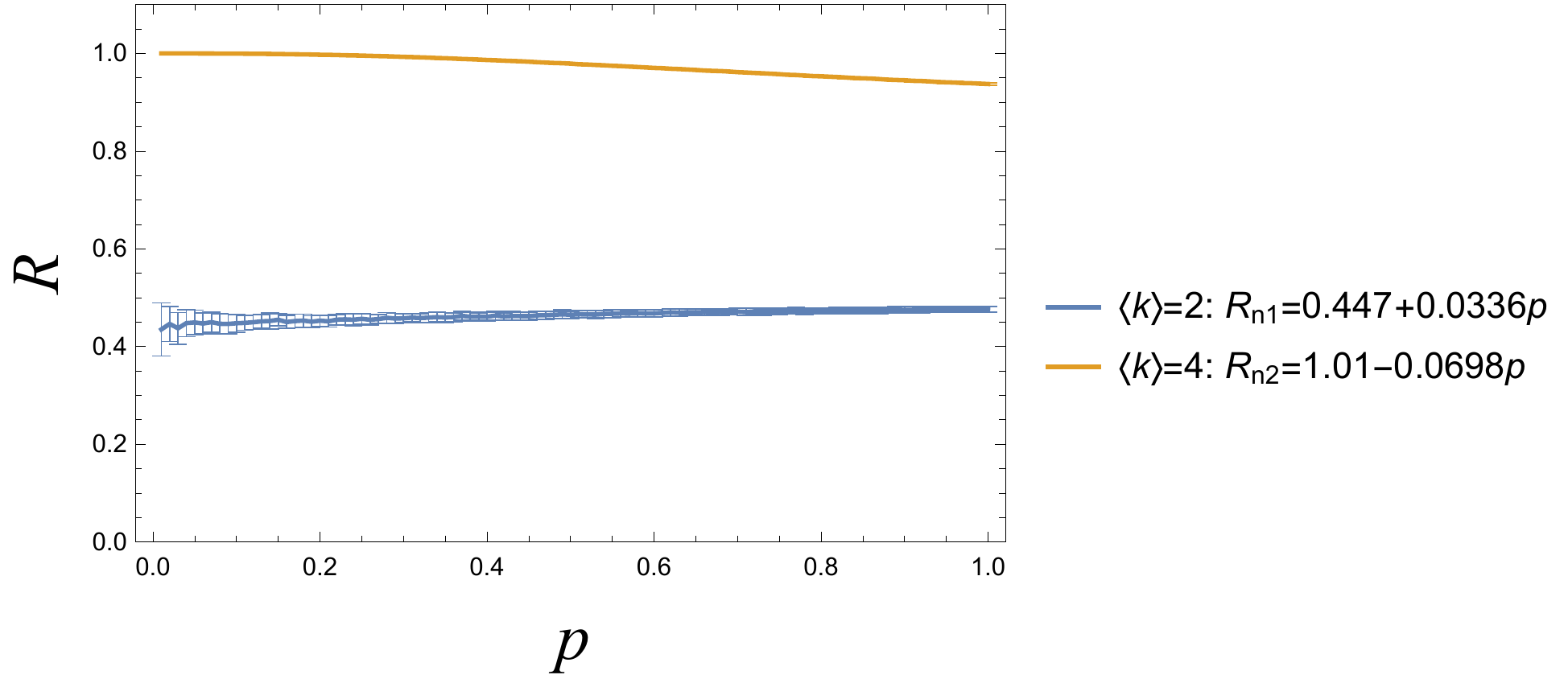}}
  \end{tabular}
  \caption{(Colour online) (a) The relation between CNR and the number of nodes in WS networks with fixed average degree 2 and rewiring probability 0.1 and 0.2. The lines are fittings. (b) The relation between CNR and rewiring probability of WS networks with 10000 nodes. Here, $R$ represents the CNR, $p$ represents the rewiring probability of networks, and $N$ represents the number of nodes in networks.}
\label{fig:CNRws}
\end{figure}

The simulation results of WS networks are shown in Fig.~\ref{fig:CNRws}. Firstly, the average degree is fixed at 2, and we study the change of CNR with network size at rewiring probabilities of 0.1 and 0.2, respectively. Since the WS networks are generated from regular networks (with mean degree of 2), their CNR have higher uncertainty in the rewiring process. The CNR will stabilize with increased network size. And the CNR is 0.45 if the average degree is 2 and 1 if the average degree is greater than 4 with low rewiring probability. But they show different tendencies when the rewiring probability is increased. The CNR of WS networks with average degree 4 will decrease with the rewiring probability and the CNR of WS networks with average degree 2 will increase slowly with rewiring probability. This is due to the fact that WS networks with high rewiring probability tend to be like ER networks. It is then not strange that the CNR's of the two are close to one another.

In the modeling of BA networks, if exactly one edge and one node are added at every step, the CNR will be 0 in this network, since this network is a tree network. Otherwise, the CNR will be 1 if more edges than one are added since every node constitutes a cycle when added.

\begin{figure}[htp]
\centering
  \begin{tabular}{cc}
  \subfigure[The CNR and TCR in ER.]{\label{fig:cnotcoer}
    \includegraphics[width=.45\textwidth]{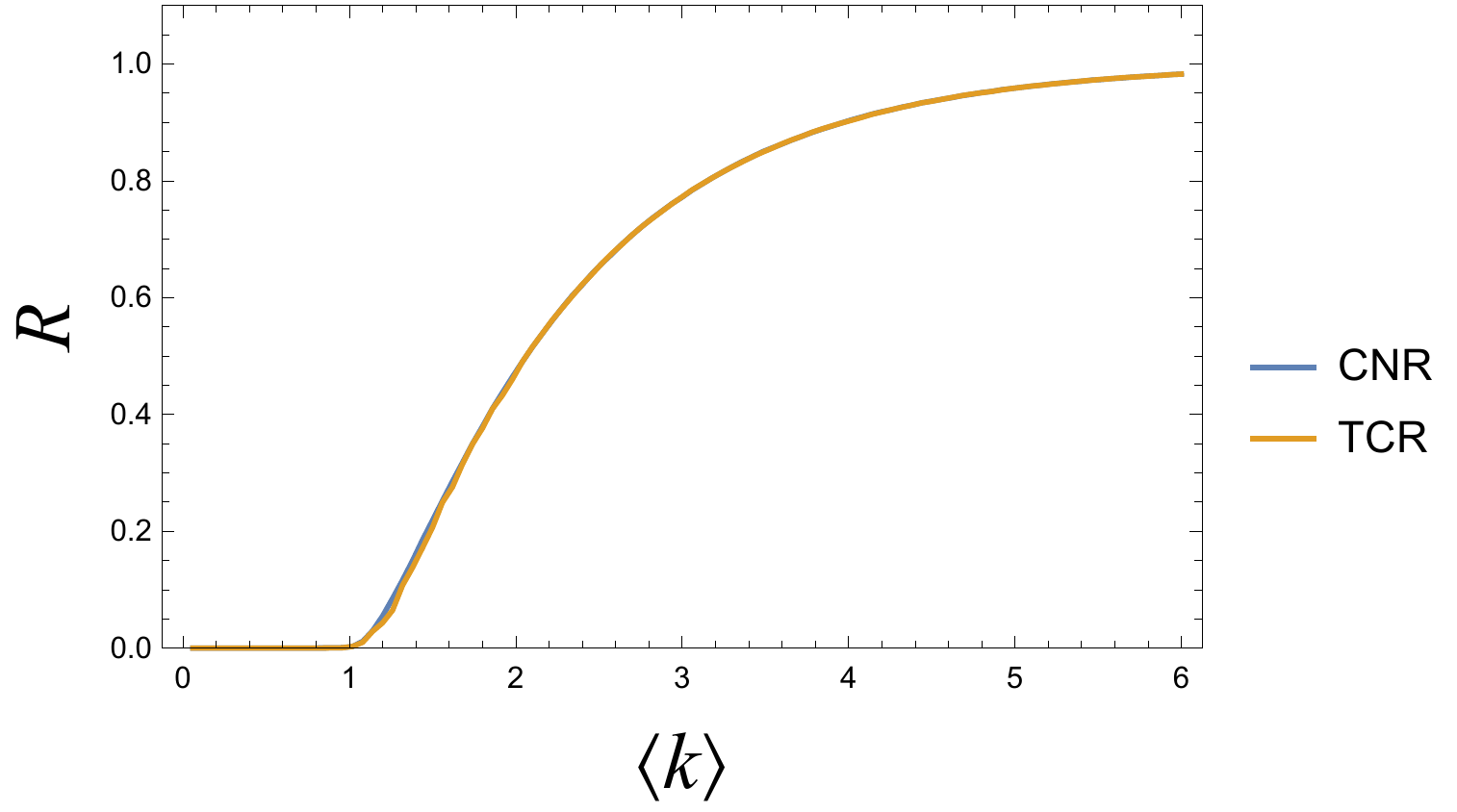}}&
    \subfigure[The CNR and TCR in WS.]{\label{fig:cnotcows}
    \includegraphics[width=.45\textwidth]{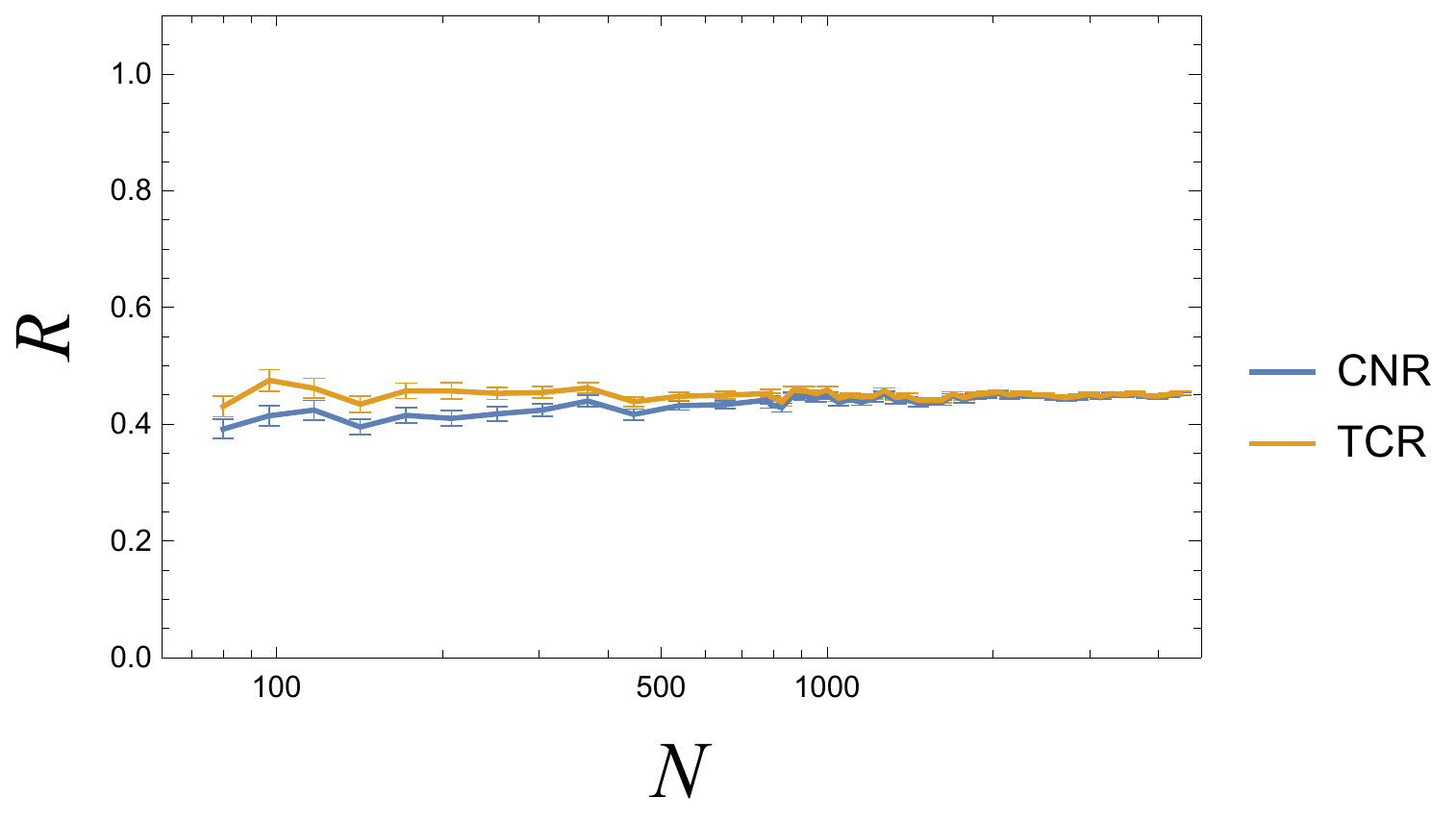}}
  \end{tabular}

\caption{(Colour online) (a) The CNR and TCR in ER, where the horizontal axis is the average degree of networks, and the vertical axis is the ratio of CNR or TCR. (b) The CNR and TCR in WS networks whose average degree is 2 and rewiring probability is 0.1. The horizontal axis is the size of networks, and the vertical axis is the ratio of CNR or TCR.}
\label{fig:cnotcomodel}
\end{figure}

\begin{figure}[htp]
\centering

  \begin{tabular}{cc}
  \subfigure[One network to show the difference between CNR and TCR.]{\label{fig:cnotco:a}
    \includegraphics[width=.45\textwidth]{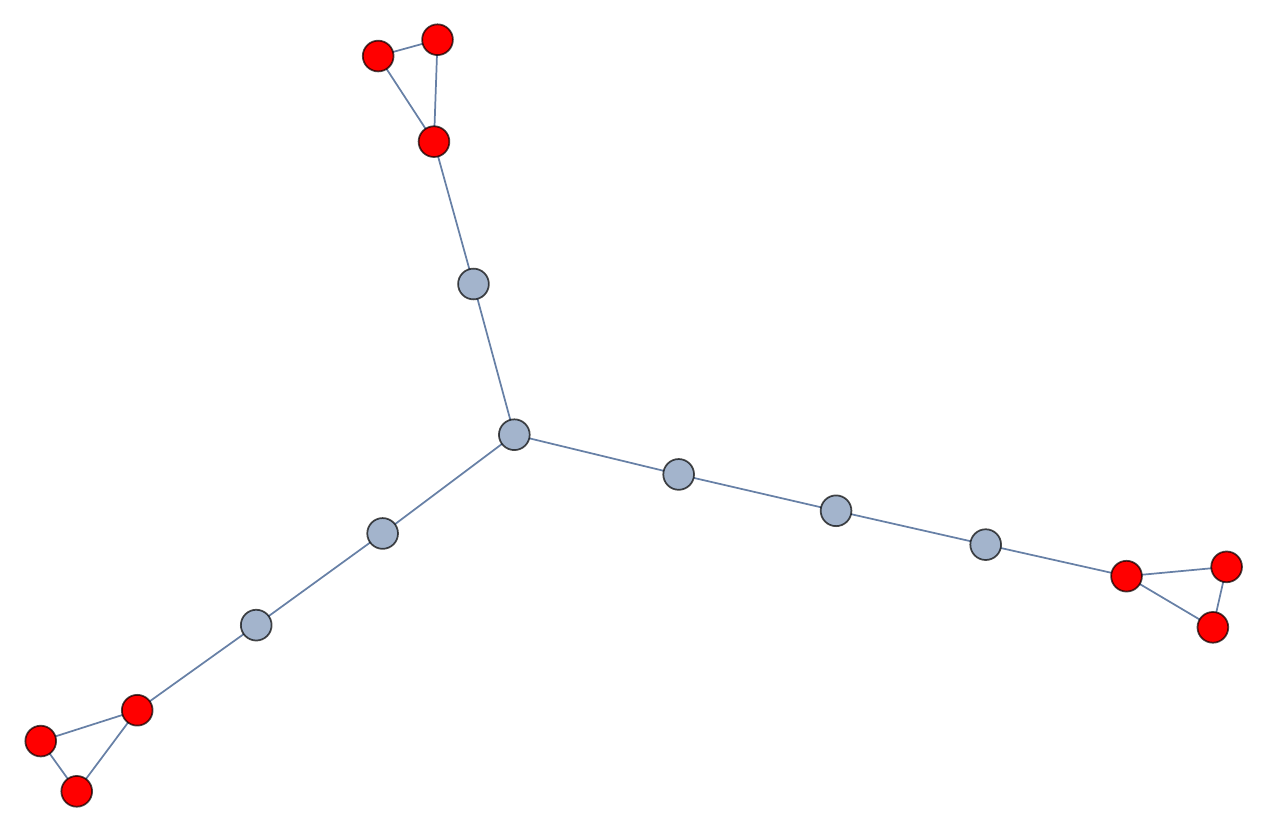}}&
    \subfigure[Fungal growth network.]{\label{fig:cnotco:b}
    \includegraphics[width=.45\textwidth]{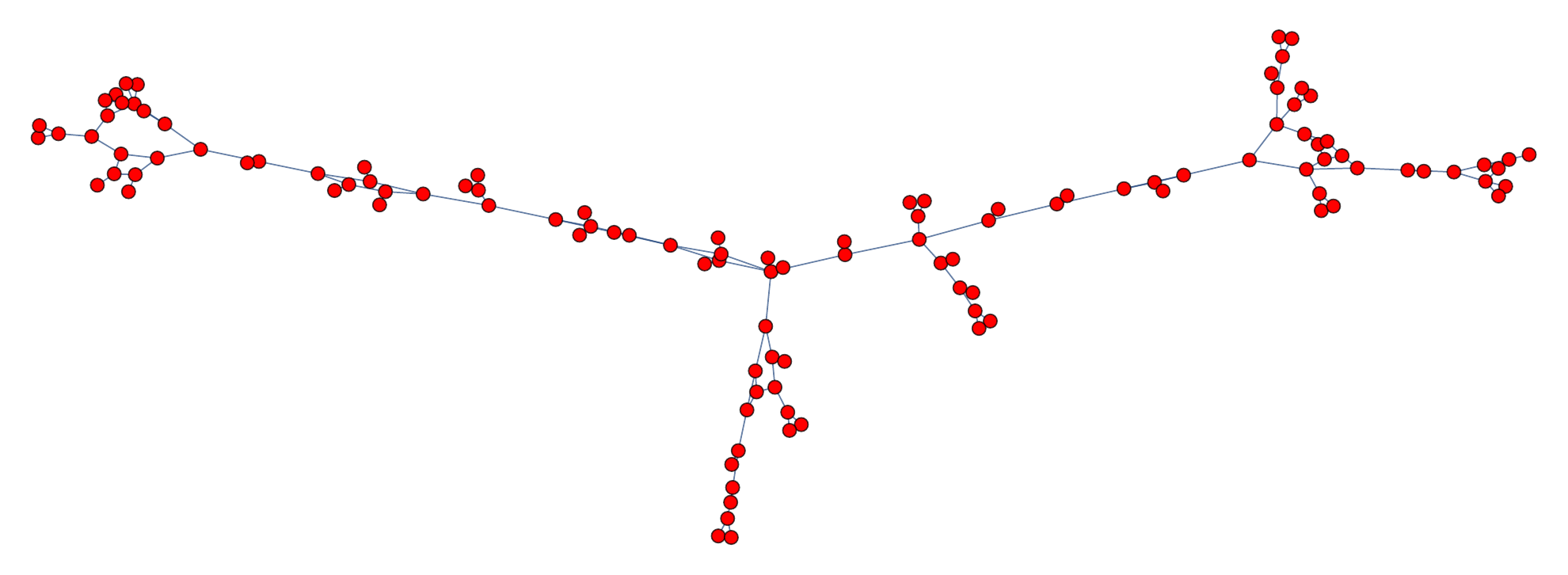}}
  \end{tabular}
\caption{(Colour online) To show the difference between cycle nodes and two-core nodes. The red nodes are cycle nodes, and the rest nodes are two-core nodes in (a), and one sample of fungal growth network (b).}
\label{fig:cnotco}
\end{figure}

To compare the difference between cycle node and two-core node, we study the CNR and two-core-ratio (TCR) in both ER networks and WS networks. CNR and TCR are very similar in ER networks, but much different in WS networks (shown in Fig.~\ref{fig:cnotcomodel}). As shown in Fig.~\ref{fig:cyclenodes}, cycle connect nodes (node 7) is the key factor that leads to the difference in cycle node and two-core node. In WS networks, cycle connect nodes are easy to generate as WS networks have a single line (the line constituted by nodes and edges one by one), the nodes contained in this kind of structure are often seen as two-core nodes but unlikely to be cycle nodes. For example, Fig.~\ref{fig:cnotco:a} shows the difference between cycle nodes and two-core nodes. This cycle connect nodes appears frequently in fungal growth networks (shown in Fig.~\ref{fig:cnotco:b}).

\begin{figure}[htp]
\centering
\includegraphics[width=.7\textwidth]{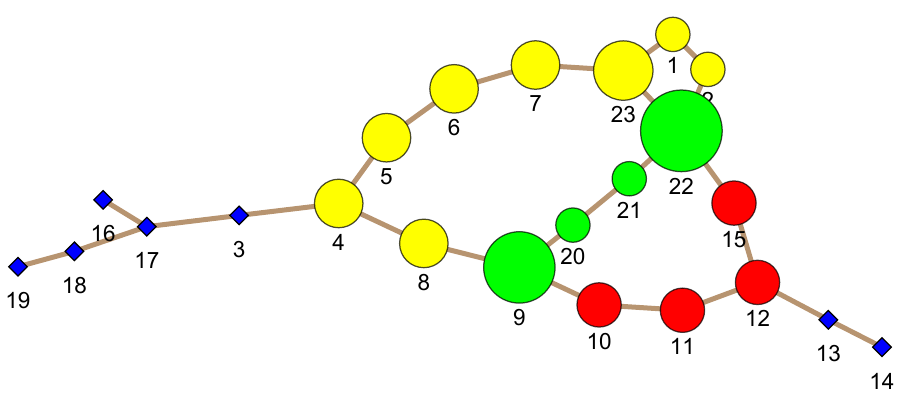}
\caption{(Colour online) The giant component of WS network with 27 nodes and 28 edges, the size of nodes represents the cycle centrality of this node. Node 22 can influence all circle shape nodes, its cycle centrality is 16, and node 12 influence the red nodes, node 23 influence the yellow nodes, which cycle centrality are 4 and 8 respectively.}
\label{fig:cyclecentrality}
\end{figure}

The node centrality is introduced to describe the importance of nodes in complex networks. The commonly used centrality measures are degree centrality, betweenness centrality \cite{freeman1977set}, PageRank centrality \cite{google2011facts}, closeness centrality \cite{bavelas1950communication}, etc. We have defined the cycle centrality of node i as the number of nodes which are transformed from cycle nodes to none-cycle nodes with node i being deleted. For example, in the below Fig.~\ref{fig:cyclecentrality}, node 22 has the largest cycle centrality, that is 16, in this network, and its deletion will transform cycle nodes to none-cycle nodes. Therefore, cycle centrality can represent the over-distance connection capability of the node. Because if you delete this node, the shortest-path lengths of its neighboring nodes will greatly increase. The node which has large cycle centrality can decrease the average shortest-path length of network.
\subsection{\label{analysisresolution} The analytical result of CNR in ER networks}
\begin{figure}[htp]
\centering
\includegraphics[width=.7\textwidth]{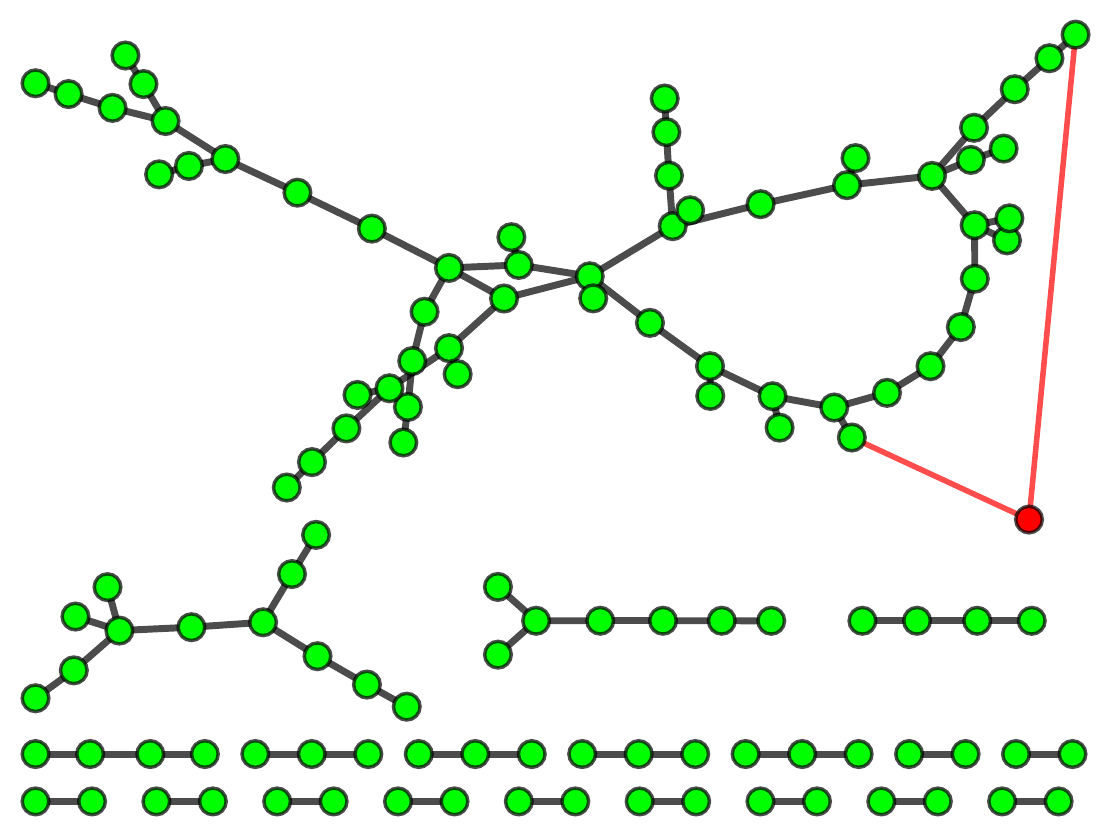}
\caption{(Colour online) The red node is cycle node if it connects to one component with two edges.}
\label{fig:componentcycle}
\end{figure}

In this part, the analytical solution of CNR in ER networks is given. Generally speaking, one node could be a cycle node if it is connected to two nodes that belong to one component in a network (The two nodes could contact through this network, as shown in Fig.~\ref{fig:componentcycle}). Suppose there are $c$ components in ER-network $G$, and $L_r$ is the number of nodes in component $r$. The number of all contact pairs is

\begin{figure}[htp]
\centering
\includegraphics[width=.7\textwidth]{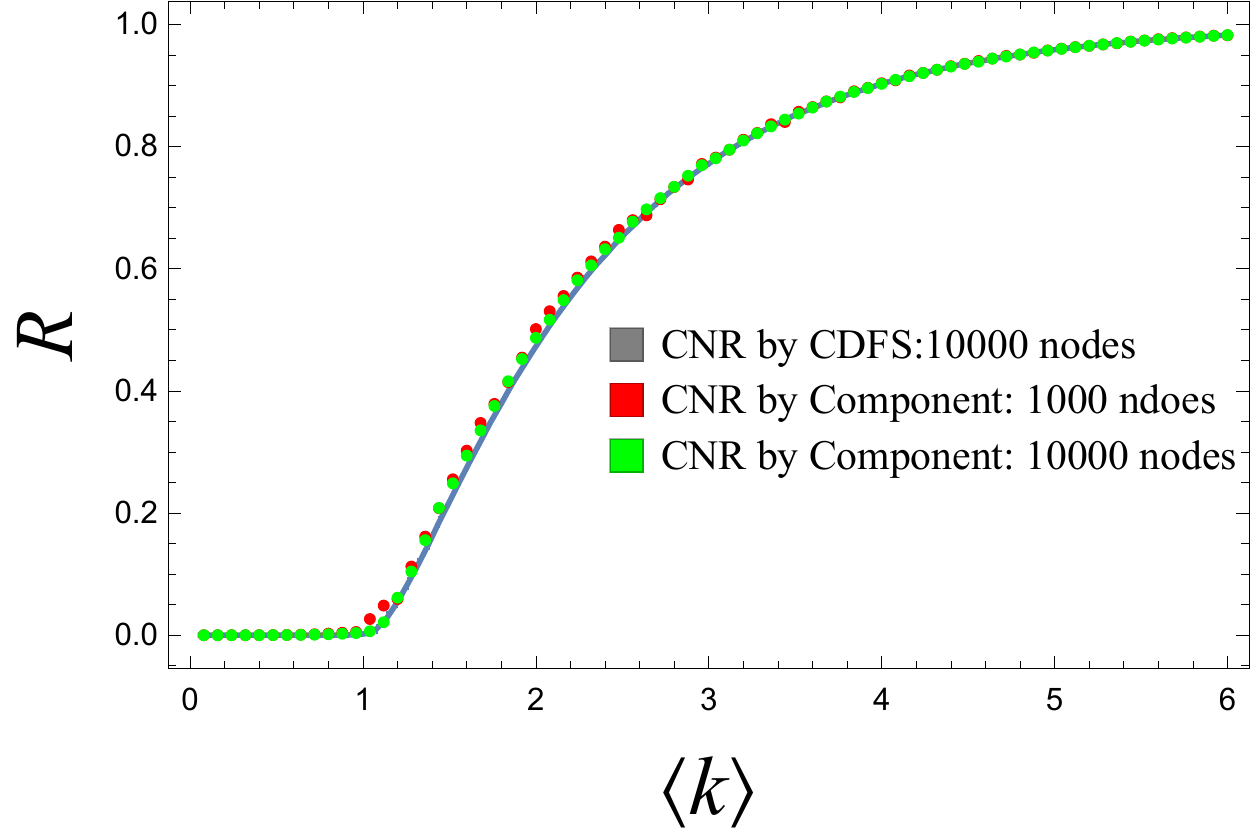}
\caption{(Colour online) The comparison between CNR in ER network with 10000 nodes (gray points), CNR calculated by component in ER-network with 1000 nodes (red points), and CNR calculated by component in ER-network with 10000 nodes (green points).}
\label{fig:copmparecomponent}
\end{figure}

\begin{equation}
n_c=\sum_{r=1}^c\frac{ L_r(L_r-1)}{2}.
\end{equation}
Then the probability that nodes with degree 2 are cycle nodes is
\begin{equation}
\label{eq:connectpair}
p_l=2n_c/n(n-1)\approx\sum_{r=1}^c\frac{L_r^2}{n^2},
\end{equation}
\noindent
where $n$ is the nodes number of the ER network. Otherwise the probability of finding the node with degree $k$ in ER networks is $P(k)$ \cite{Newman2001Random}:
\begin{equation}
\label{eq:pk}
P(k)={n \choose k}p^k(1-p)^{n-k},
\end{equation}
\noindent
where $p$ is the connecting probability of different nodes in ER-networks. And the probability $p_o$ of nodes, with different degrees, being a cycle node in ER networks is shown in Tab.~\ref{Tab:ercomponent}. The summation of the probabilities over all nodes in ER-networks gives the CNR
\begin{equation}
\label{eq:op}
R=1-P(0)-\sum_{k=1}^{n}P(k)(1-p_l)^{\frac{k(k-1)}{2}}.
\end{equation}
Then we have the simulation of CNR in ER networks by Eq.~\ref{eq:op} which is shown in Fig.~\ref{fig:copmparecomponent}. The CNR calculated by the network components is fairly consistent with the simulation result.
\begin{table}[!htp]
\centering
\caption{The probability that nodes, with different degrees, are cycle nodes. Here $k$ is the degree, $p_o$ is the probability of the node as cycle node, and $P(n)$ is the probability of nodes with degree $n$.}
\label{Tab:ercomponent}
\begin{tabular}{cc}
\hline
$k$&$p_o$\\
\hline
2&$P(2)p_l$\\
3&$P(3)(1-(1-p_l)^3)$\\
4&$P(4)(1-(1-p_l)^6)$\\
\vdots & \vdots \\
$n$&$P(n)(1-(1-p_l)^{\frac{n(n-1)}{2}})$\\
\hline
\end{tabular}
\end{table}

\begin{figure}[htp]
\centering
\includegraphics[width=.7\textwidth]{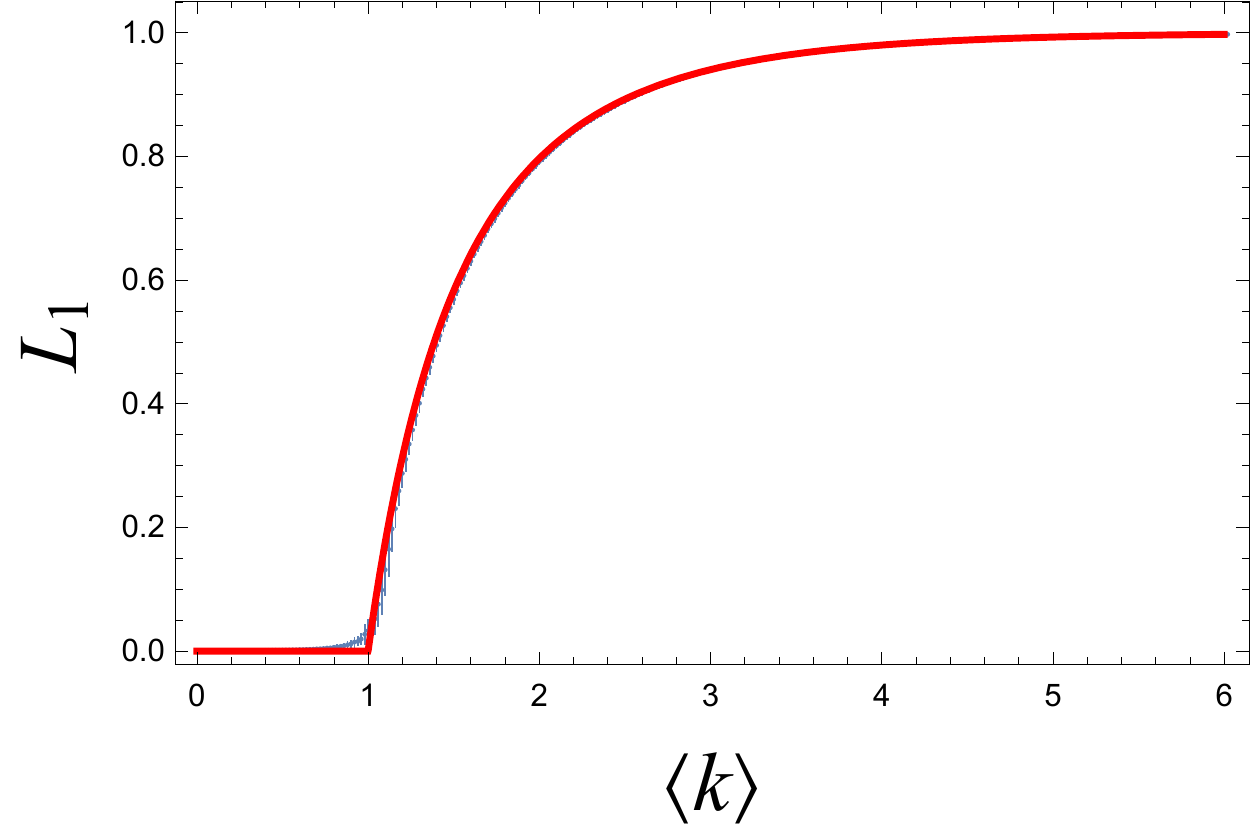}
\caption{(Colour online) The comparison between simulation result (gray points) and analytical result (red line) of normalization size of giant component $L_1$ in ER-network with 10000 nodes. They collapse onto each other very well except the transition point of the average degree being $1$ (Since the finite size effect).}
\label{fig:giantanalytical10000}
\end{figure}

\begin{figure}[htp]
\centering
\includegraphics[width=.7\textwidth]{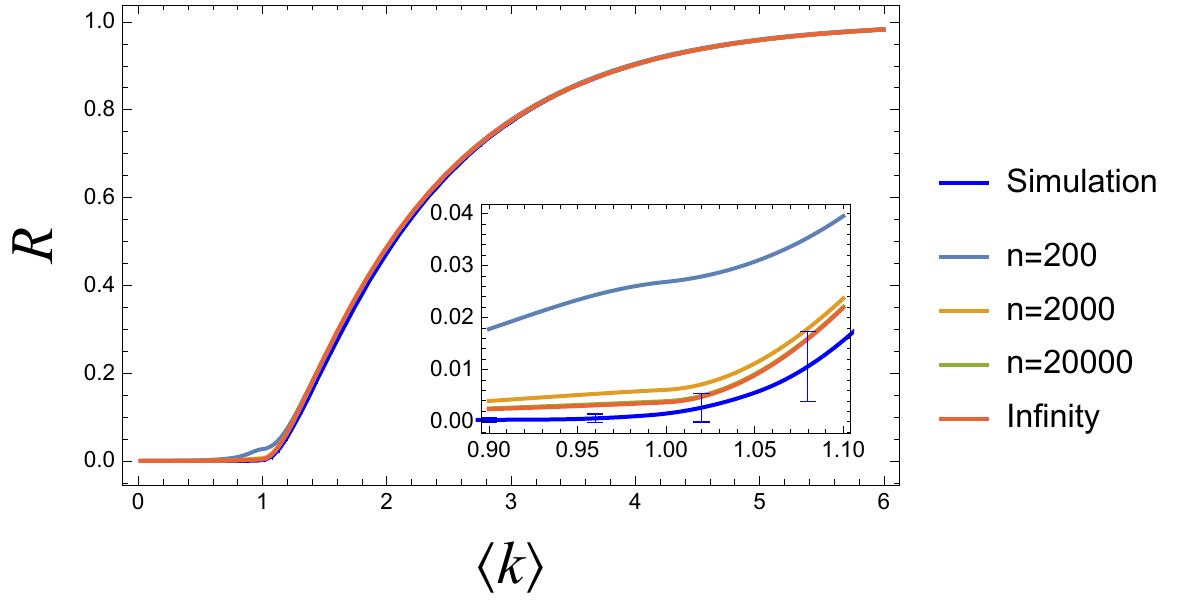}
\caption{(Colour online) The comparison between simulation results (green line contains 20000 nodes.) and analytical results of normalized size of giant component $L_1$ in ER-network with 200, 2000, 20000, and infinite nodes. The curves collapse onto each other very well except at the transition point of the average degree being $1$ (due to the finite-size effect).}
\label{fig:ER10000CNRanalytical}
\end{figure}

Consider the ER-network $G(n,p)$ with size $n$ and connecting probability $p$, the number of nodes on tree components of $G(n,p)$ ($n\rightarrow \infty$) is (see in Red.~\cite{bollobas2001random})
{\setlength\arraycolsep{2pt}
\begin{eqnarray}
\label{eq:ktr}
E\left(\sum_{l=1}^n lT_l\right)&=&\frac{n}{\langle k\rangle}\sum_{l=1}^{\infty}\frac{l^{l-1}}{l!}(\langle k\rangle e^{-\langle k\rangle})^l+O(1)\nonumber\\
&=&nt(\langle k\rangle),
\end{eqnarray}}
where $T_l$ is the number of components in $G(n,p)$ that are trees of size $l$, and $\langle k\rangle$ is the average degree of $G(n,p)$. $t(\langle k\rangle)$ is the probability of a node belonging to the tree component. By the manipulations of Eq.~\ref{eq:ktr}, $s=s(\langle k\rangle)=\langle k\rangle t(\langle k\rangle)$ is the only solution of
\begin{equation}
se^{-s}=\langle k\rangle e^{-\langle k\rangle}
\end{equation}
in the range $0<s<1$. So
\begin{equation}
t(\langle k\rangle)=\frac{s(\langle k\rangle)}{\langle k\rangle}=1 \qquad \textrm{for} \quad 0<\langle k\rangle<1.
\end{equation}
This result means that all nodes on the tree components, and the number of nodes on cycles is 0 for $0<\langle k\rangle<1$. In the range $\langle k\rangle>1$, $G(n,p)$ is the union of the giant component $L_1(G(n,p))$, the small unicyclic components and the small tree components. There are at most $\omega(n)$ vertices on the unicyclic components \cite{bollobas2001random}. The size of the giant component satisfies (see in Ref.~\cite{bollobas2001random})
\begin{equation}
|L_1(G(n,p))-\{1-t(\langle k\rangle)\}n|\leq \omega(n)n^{1/2}.
\end{equation}
From Ref.~\cite{bollobas2001random} we get
\begin{equation}
 \omega(n)\sim \frac{1}{2}\sum_{l=3}^{\infty}(\langle k\rangle e^{-\langle k\rangle})^l\sum_{j=0}^{l-3}l^j/j!=\mu(\langle k\rangle),
\end{equation}
and since
{\setlength\arraycolsep{2pt}
\begin{eqnarray}
\lim_{l\to\infty}(\langle k\rangle e^{-\langle k\rangle})^l\sum_{j=0}^{l-3}l^j/j!&=&(\langle k\rangle e^{-\langle k\rangle})^le^l\nonumber\\
&=&(\langle k\rangle e^{1-\langle k\rangle})^l\nonumber\\
&=&o(l^{-M}),
\end{eqnarray}}
for every $M>0$, since $xe^{1-x}\leq 1$ for $x>0$. Thus $\omega(n)=O(1)$, and
\begin{equation}
\label{eq:treelargecycle}
\omega(n)\ll\omega(n)n^{1/2}\ll (1-t(\langle k\rangle))n \quad \textrm{for} \quad n\rightarrow \infty.
\end{equation}
Then we conclude that (according to \cite{erdHos1960evolution,erdHos1961strength})
\begin{equation}
L_1(G(n,p))\approx (1-t(\langle k\rangle))n  \qquad \textrm{for} \quad n\rightarrow \infty.
\end{equation}
This result is shown in Fig.~\ref{fig:giantanalytical10000}, which is consistent with the simulation result. By Eq.~\ref{eq:connectpair}, $p_l$ can be calculated as:
{\setlength\arraycolsep{2pt}
\begin{eqnarray}
\label{eq:connectpairayalytical}
p_l&\approx &\sum_{r=1}^c\frac{L_r^2}{n^2}\nonumber\\
&=&\frac{1}{n^2}(L_1^2+\sum_{r=2}^cL_r^2)\nonumber\\
&\approx&(1-t(\langle k\rangle))^2+\frac{1}{n^2}\sum_{l=2}^nl^2T_l\nonumber\\
&=&(1-\frac{1}{\langle k\rangle}\sum_{l=1}^{\infty}\frac{l^{l-1}}{l!}(\langle k\rangle e^{-\langle k\rangle})^l)^2\nonumber\\
&+&\frac{1}{\langle k\rangle n}\sum_{l=1}^{\infty}\frac{l^l}{l!}(\langle k\rangle e^{-\langle k\rangle})^l,
\end{eqnarray}}
where Eq.~\ref{eq:treelargecycle} is used, and the third term could be omitted if $n\rightarrow \infty$. Combining Eqs.~\ref{eq:pk}, ~\ref{eq:op}, ~\ref{eq:connectpairayalytical} and $\langle k\rangle=np$, the CNR could be written as:
{\setlength\arraycolsep{2pt}
\begin{eqnarray}
\label{eq:Oanalytical}
R&=&1-(1-p)^n\nonumber\\
 &-&\sum_{k=1}^{n}{n \choose k}p^k(1-p)^{n-k}(1-(1-\frac{1}{np}\sum_{l=1}^{\infty}\frac{l^{l-1}}{l!}(np e^{-np})^l)^2)^{\frac{k(k-1)}{2}}.
\end{eqnarray}}
Since $(1-p_l)^{\frac{k(k-1)}{2}}$ decays rapidly with $k$, so we calculate it for $k<5$, which is shown in Fig.~\ref{fig:ER10000CNRanalytical}. The analytical result is in accordance with the simulation result. And the giant component, which appears when $\langle k\rangle=1$, leads to the turning point of CNR. The analytical result is close to the analytical result of TCR which has been observed in Ref.~\cite{zhao2013inducing}. These two analytical results have different analytic forms which are caused by different methods, even though they have very similar results in ER networks. Besides, our method is easier to understand and calculate.

\section{\label{sec:CNRinclassification}The CNR in real networks and its application in network classification}
\subsection{CNR in real networks, and used in conjunction with improved spectral coarse-graining}

To study the properties of CNR in real networks, we employ the data samples from network repository \cite{nr}, Stanford Large Network Dataset \cite{snapnets},  Pajek datasets \cite{batagelj2006pajek}, and some classic complex networks \cite{kolaczyk2009analysis,newman2001structure,newman2001scientific,Sucharita2007The,lusseau2003emergent,girvan2002community,zachary1977information}. $30$ unweighted and undirected complex networks were chosen randomly. The size of these networks is between $34$ and $65533$, and their edges range from $78$ to $51971$.

\begin{figure}[htbp]
\centering
    \includegraphics[width=.7\textwidth]{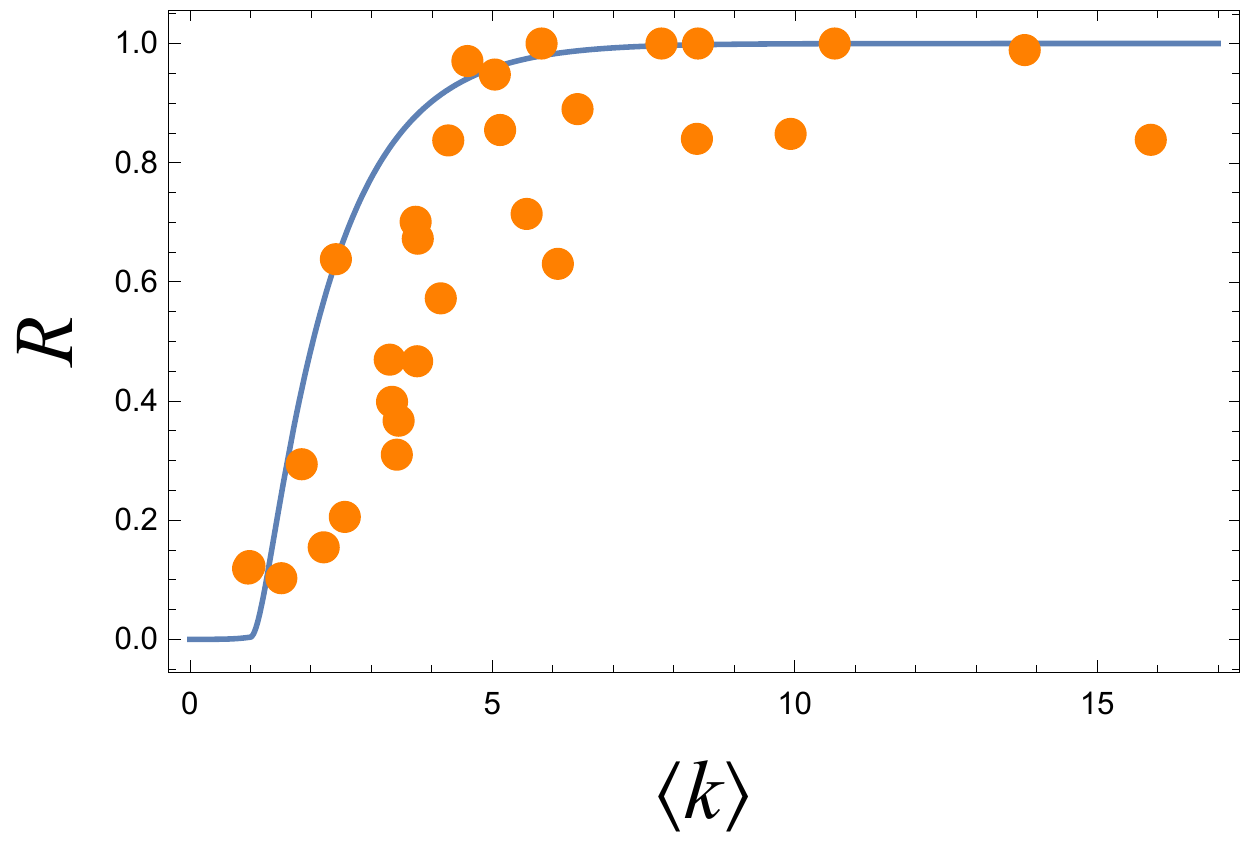}
  \caption{(Colour online) The CNR of some real networks (green points), compared with ER networks (red line), where $R$ is the value of CNR, and $k$ is the average degree of networks. The 30 networks were randomly chosen from 4 datasets. The green datasets have a similar trend with that of the red line, but most green points are below the red line.}
  \label{fig:LVOinreal}
\end{figure}

Their CNR and average $k$ are calculated and compared with ER network, as shown in Fig.~\ref{fig:LVOinreal}. The CNR of real networks increases with the increase of average degree, and has a similar trend with that of the ER networks. But it is obvious that the CNR of real networks is lower than that of the ER networks and much lower than that of the BA networks and WS networks (the value of CNR is $1$ in BA and WS networks when the degree is large than 4). So, we argue that the networks generated by basic network models cannot produce the exact CNR properties. The real networks are much more tree-like compared with the network models.

\begin{figure}[htp]
\centering
  \begin{tabular}{cc}
  \subfigure[One diag network.]{\label{fig:coarse:a}
    \includegraphics[width=.45\textwidth]{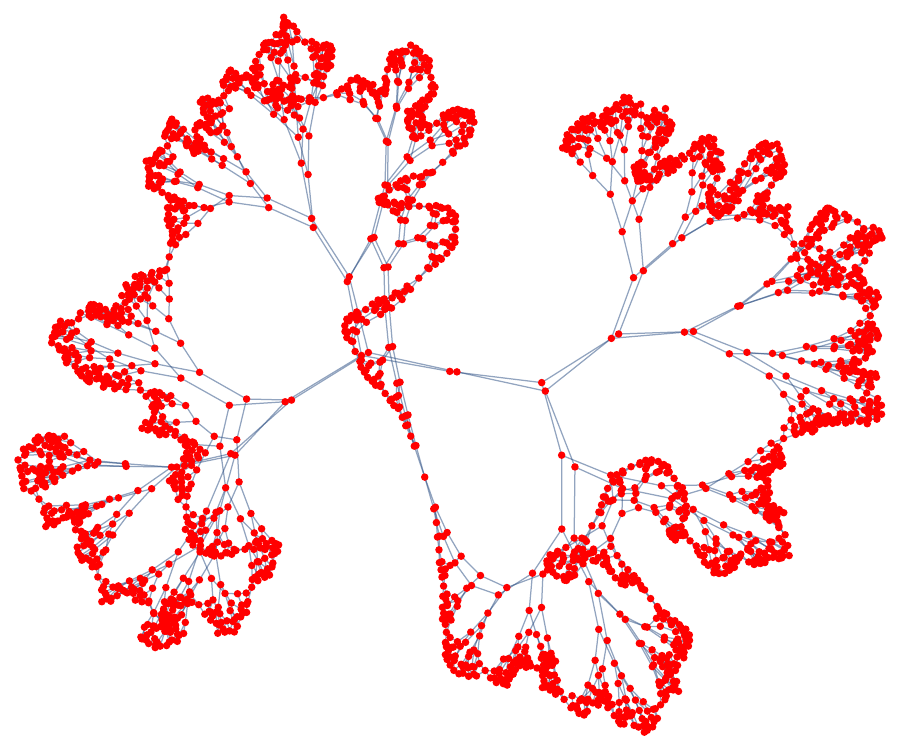}}&
    \subfigure[One chemistry molecules network.]{\label{fig:coarse:b}
    \includegraphics[width=.45\textwidth]{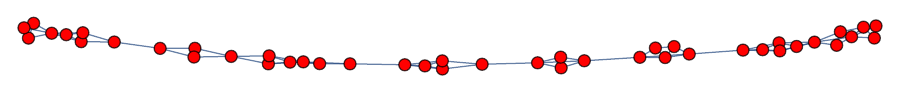}}\\
      \subfigure[The coarsely grained diag network.]{\label{fig:coarse:c}
    \includegraphics[width=.45\textwidth]{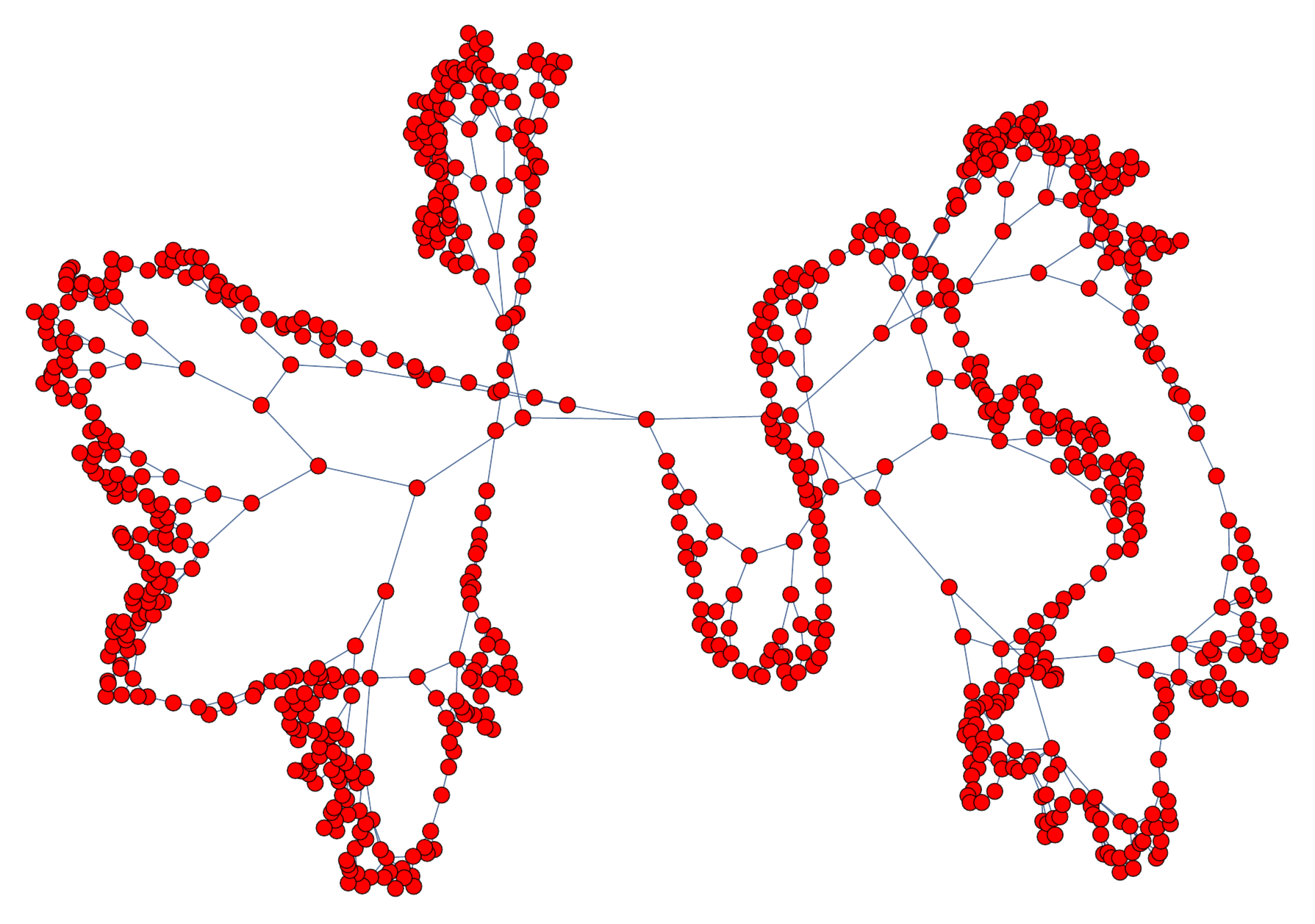}}&
    \subfigure[The coarsely grained chemistry molecules network.]{\label{fig:coarse:d}
    \includegraphics[width=.45\textwidth]{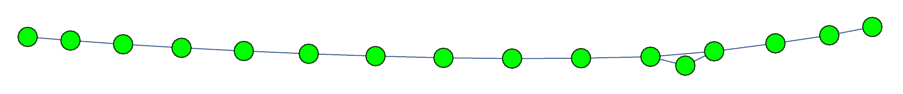}}
  \end{tabular}
\caption{(Colour online) The coarse grain result ((c) and (d))of networks with high average degree which are a diag network (a) and one chemistry network (b).}
\label{fig:coarse}
\end{figure}

In real networks, we find that most networks with high average degree have CNR close to 1. But their macroscopic properties do not look like cycles if we zoom out the networks. So if we want to measure the CNR of macro states of these graphs, we should combine the improved spectral coarse-graining (ISCG) \cite{zhou2017improved}. The ISCG is a new method to coarse-grain the graph based on the spectral coarse-graining (SCG) \cite{gfeller2007spectral}. Compared to the SCG method, the ISCG algorithm has many advantages such as smaller errors, better effects, and greatly reduced computational complexity. For the two networks in Fig.~\ref{fig:coarse}, both of their CNR are 1, but it is obvious that the right one is more like cycle and the left one is more close to a line. Coarse-graining these two networks by ISCG, we got the corresponding new networks under them (the number of nodes is reduced by two-thirds). The new CNR of these two networks is 1 and 0.1875, respectively. In the transport and flow networks, the macro-structure is important than the micro-structure. In this case, calculation of the CNR of the coarse-graining networks is useful and feasible.

\subsection{The application of CNR in network classification}
\begin{figure}[htbp]
\centering
    \includegraphics[width=.7\textwidth]{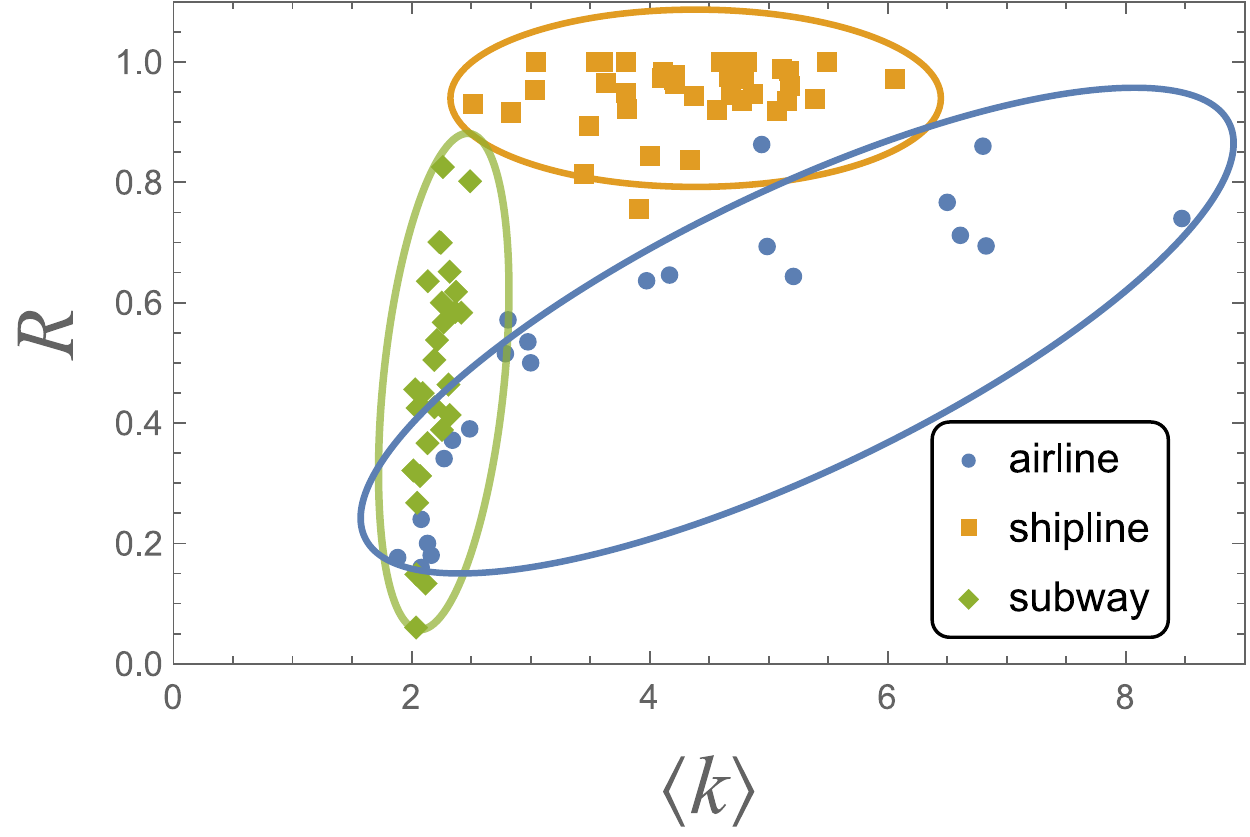}
  \caption{(Colour online) The CNR of three transport network sets which are airline (blue points), ship line (orange points), and subway (green points). Every point represents one transport company or city. The points belong to same kind of network tend to together.}
  \label{fig:LVOintransport}
\end{figure}

Then we study the transport networks, and find that different kinds of transport networks tend to form different zones which barely overlap with each other (shown in Fig.~\ref{fig:LVOintransport}). That is to say, they can be classified by different zones. To be more compelling, we also included the network set of public transportation in different cities and fungal growth network for classification. The datasets for classification contain 28 subway networks of different cities, 41 marine transport networks of different companies, 21 airline networks of different companies, 405 cities' public transportation networks of different cities in China, and 270 fungal growth networks. All these datasets can be downloaded from Ref.~\cite{CNR2019}. The basic properties of these networks are shown in Table.~\ref{Tab:netset}.

\begin{table}[!htp]
\centering
\caption{The size $n$, number of edges $m$, mean degree $\langle k\rangle$, and global clustering coefficient $C$ of 5 different kinds of networks.}
\label{Tab:netset}
\begin{tabular}{ccccc}
\hline
Network&$n$&$m$& $\langle k\rangle$&$C$\\
\hline
Subway networks&68$\sim$595&69$\sim$741&2.015$\sim$2.491&0$\sim$0.0319\\
Marine transport networks&12$\sim$356&17$\sim$844&2.512$\sim$6.056&0.109$\sim$0.519\\
Airline networks&17$\sim$123&16$\sim$521&1.882$\sim$8.472&0.0803$\sim$0.493\\
Fungal growth networks&69$\sim$2741&72$\sim$3665&1.607$\sim$3.254&0.00595$\sim$0.207\\
Public transportation networks&8$\sim$13133&8$\sim$19905&2$\sim$4.620&0$\sim$0.152\\
\hline
\end{tabular}
\end{table}
These five datasets are classified by employing the machine learning metric of K-Nearest Neighbor (KNN) algorithm \cite{keller1985fuzzy}. The basic process of the algorithm is to calculate the distance in characteristic space and identify the category of most samples of the $K$ nearest neighbor of the new sample belongs to, then set the new sample belonging to this category. As the results of Ref.~\cite{kantarci2013classification}, we choose the four most discriminant features and CNR to be the machine learning parameters: the density $\delta$, modularity $Q$, average degree $\langle k\rangle$,  global clustering coefficient $C$, and CNR. The density $\delta(G)$ is the ratio of existing edges $m$ to the number of all possible edges in network $G$.
\begin{equation}
\delta(G)=\frac{m}{n(n-1)}.
\end{equation}
The modularity is the quantity which shows the quality of the community structure. It is the proportion of links inside the communities minus the estimated result from the corresponding null model. The global clustering coefficient refers to the proportion of triangles, which measures the connection density among neighboring nodes of one node.

We classify this network set in six different approaches. In the first one, all five features are chosen, and four features out of five are chosen in the next five approaches. The classification result is shown in Tab.~\ref{Tab:machineresults}, and the accuracy of classification without the CNR is significantly lower than the rest ones with the CNR. This means that the CNR can increase the accuracy of network recognition in machine learning obviously. So we think the CNR is one important property of networks, which can not only show whether the networks tend to be tree-like or cycle-like,  but can distinguish different kinds of networks.

\begin{table}[!htp]
\centering
\caption{The classifications of datasets trained by KNN Classifier algorithm with the features of  density $\delta$, modularity $Q$, average degree $\langle k\rangle$,  global clustering coefficient $C$, and CNR $R$. }
\label{Tab:machineresults}
\begin{tabular}{ccccc}
\hline
Features&Precision&Recall&F1-score&Support\\
\hline
All five features&0.91&0.92&0.91&192\\
Without $\delta$&0.90&0.91&0.90&192\\
Without $Q$&0.89&0.90&0.89&192\\
Without $\langle k\rangle$&0.89&0.90&0.89&192\\
Without $C$&0.91&0.92&0.91&192\\
Without $R$&\textbf{0.80}&\textbf{0.80}&\textbf{0.79}&192\\
\hline
\end{tabular}
\end{table}
\section{Conclusions}
In this paper, we have proposed the concept of CNR to judge whether the network tends to be treelike or not, and an algorithm to calculate this quantity. The CNR is constant with different sized networks, and displays critical turning points in ER networks. TCR is greater than CNR in WS networks and some real networks. The cycle centrality is introduced to describe over-distance connection capability of the node among its neighboring nodes. We have provided the analytical solution of CNR in ER networks, which can explain the critical turning in ER networks by giant component. We then study the CNR in real networks, and find that its value is lower than that of the basic networks models with the same average degree. In addition, it is useful to combine the coarse-graining method to judge the cycle structure of networks. Finally, we have classified the networks of different kinds by machine learning. The CNR is found to be the most compelling feature to enhance the accuracy of classification. Therefore we may regard the CNR as one essential quantity of complex networks.
\section*{Acknowledgements}
We gratefully acknowledge Linyuan L{\"u}, Haijun Zhou, Liping Chi, and Shengfeng Deng, give us fruitful suggestions. Thanks to Yueying Zhu, Longfeng Zhao, and Jiao Gu for offer help on network data. This work was supported in part by National Natural Science Foundation of China (Grant No. 61873104, 11505071), the Programme of Introducing Talents of Discipline to Universities under Grant No. B08033, the Fundamental Research Funds for the Central Universities.

\bibliography{mybibfile}
\end{document}